\documentclass[twocolumn,aps,showpacs]{revtex4}

\usepackage{graphicx}
\usepackage{amssymb}
\usepackage[latin1]{inputenc}
\usepackage{bm}

\begin{document}

\title{Entangled electron current through finite size \\ normal-superconductor tunneling structures}

\author{E. Prada}
 \email{elsa.prada@uam.es}
 \affiliation {Departamento de F\'{\i}sica Te\'orica de la Materia Condensada, C-V, and
 Instituto Nicol\'as Cabrera,
 Universidad Aut\'onoma de Madrid, E-28049 Madrid, Spain}
\author{F. Sols}
 \email{fernando.sols@uam.es}
 \affiliation {Departamento de F\'{\i}sica Te\'orica de la Materia Condensada, C-V, and
 Instituto Nicol\'as Cabrera,
 Universidad Aut\'onoma de Madrid, E-28049 Madrid, Spain}

\begin{abstract}
We investigate theoretically the simultaneous tunneling of two
electrons from a superconductor into a normal metal at low
temperatures and voltages. Such an emission process is shown to be
equivalent to the Andreev reflection of an incident hole. We
obtain a local tunneling Hamiltonian that permits to investigate
transport through interfaces of arbitrary geometry and potential
barrier shapes. We prove that the bilinear momentum dependence of
the low-energy tunneling matrix element translates into a real
space Hamiltonian involving the normal derivatives of the electron
fields in each electrode. The angular distribution of the electron
current as it is emitted into the normal metal is analyzed for
various experimental setups. We show that, in a full
three-dimensional problem, the neglect of the momentum dependence
of tunneling causes a violation of unitarity and leads to the
wrong thermodynamic (broad interface) limit. More importantly for
current research on quantum information devices, in the case of an
interface made of two narrow tunneling contacts separated by a
distance $r$, the assumption of momentum-independent hopping
yields a nonlocally entangled electron current that decays with a
prefactor proportional to $r^{-2}$ instead of the correct
$r^{-4}$.
\end{abstract}
\pacs{74.45.+c,74.50.+r}

\maketitle

\section{Introduction}

The electric current through a biased normal-superconductor (NS)
interface has for long been the object of extensive theoretical
and experimental attention
\cite{andreev64,griffin70,blonder82,tinkham96}. Recently, new
interest in this classic problem has been spurred by the
possibility of using conventional superconductors as a natural
source of entangled electron pairs that may be injected into a
normal or ferromagnetic metal
\cite{byers95,torres99,deutscher00,recher01,falci01,melin01,
lesovik01,apinyan02,melin02,feinberg02,recher02,chtc02,feinberg03,recher03}
and eventually used for quantum communication purposes. Clearly,
the efficient and controlled emission of electron singlet pairs
into normal metals or semiconductor nanostructures requires a
deeper understanding of the underlying transport problem than has
so far been necessary. In particular, it is of interest to
investigate how the entangled electron current depends on various
parameters such as the shape and size of the NS interface as well
as the potential barrier profile experienced by the tunneling
electrons. A preliminary focus on tunneling interfaces seems
adequate, both because such interfaces are amenable to a simpler
theoretical study and because the low electric currents which they
typically involve will facilitate the control of individual
electron pairs.

In the light of this new motivation, which shifts the attention
onto the fate of the emitted electron pairs, it seems that the
picture of Andreev reflection, which so far has provided an
efficient book-keeping procedure, has reached one of its possible
limits. When dealing with finite size tunneling contacts, the
Hamiltonian approach is more convenient than the calculation of
the scattering wave functions, since it does not require to solve
the diffraction problem to find the conductance. Moreover, it is
hard to see how problems such as the loss of nonlocal spin
correlations among distant electrons emitted from a common
superconducting source can be analyzed in terms of Andreev
reflected holes in a way that is both practical and respectful to
causality. While an Andreev description may still be practical in
situations involving multiple electron-hole conversion, the fate
of the quasiparticles in the outgoing scattering channels will
have to be investigated in terms of a two-electron (or two-hole)
picture if one is interested in studying nonlocal correlations in
real time.

Recently, several authors
\cite{deutscher00,recher01,falci01,apinyan02,melin02,feinberg02,
recher02,feinberg03,recher03} have addressed the emission of
electron pairs through two distant contacts in a language which
explicitly deals with electrons above the normal Fermi level. Here
we investigate the emission of electron pairs from a
superconductor into a normal metal through tunneling interfaces of
different geometrical shapes and potential barriers. With this
goal in mind, we devote section II to rigorously establish the
equivalence between the pictures of two-electron emission and
Andreev reflection of an incident hole. We argue extensively that
each picture reflects a different choice of chemical potential for
the normal metal, a point also noted in Ref. \cite{samu03} . After
a precise formulation of the problem in section III, we derive a
real space tunneling Hamiltonian in section IV that accounts for
the fact that electrons with different perpendicular energy are
transmitted with different probability through the interface. In
section V, we study the structure of the perturbative calculation
that, for vanishing temperatures and voltages, will yield the
electron current to lowest order in the tunneling Hamiltonian.
Section VI concerns itself with the angular dependence of the
current through a broad NS interface, providing the connection
with calculations based on the standard quasiparticle scattering
picture \cite{kupka94,kupka97}. In section VII we investigate the
tunneling current through a circular NS interface of arbitrary
radius, paying attention not only to the total value of the
current but also to its angular distribution and to the underlying
two-electron angular correlations. We also investigate how the
thermodynamic limit is achieved for broad interfaces. Section VIII
deals with the electron current through an interface made of two
distant small holes, focussing on the distance dependence of the
contribution stemming from nonlocally entangled electrons leaving
through different holes. In section IX we investigate the commonly
used energy-independent hopping model and prove that it violates
unitarity, leads to a divergent thermodynamic limit, and yields
the wrong distance dependence for the current contribution coming
from nonlocally entangled electrons. A concluding summary is
provided in section X.

\section{Two-electron emission vs. Andreev reflection}

\begin{figure}
\includegraphics[width=8.5cm]{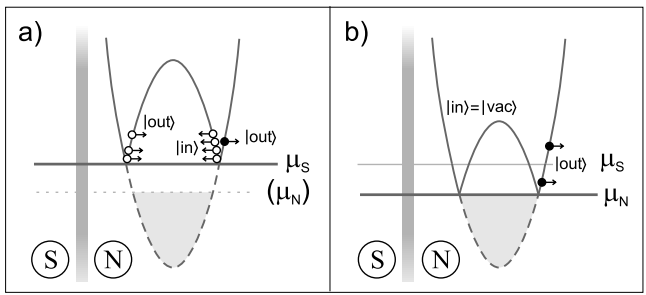} \caption{Hole Andreev
reflection vs two-electron emission: (a) When $\mu_S$ is used as
the reference chemical potential both in S and N, a typical
scattering process at the NS tunneling interface for $eV\equiv
\mu_S-\mu_N>0$ is that of incoming holes with energies between
$\mu_S$ and $\mu_S+eV$ that are most often normally reflected and
only rarely Andreev reflected. (b) If, alternatively, $\mu_N$ is
used to define quasiparticles in N, the many incoming or outgoing
holes are viewed as a vacuum of quasiparticles. The outgoing
electron generated in a rare Andreev reflection event appears now
as a spontaneously emitted electron above $\mu_S$. Such an event
causes an outgoing hole state to be empty. This is now perceived
as the emission of a second electron with energy between $\mu_N$
and $\mu_S$. Tracking the spin and the momentum component parallel
to the interface leads to the picture of two electrons emitted in
a spin singlet state with opposite parallel
momenta.}\label{2e-vs-Andreev}
\end{figure}

In a biased normal-superconductor tunneling interface in which
e.g. the superconductor chemical potential is the greatest, one
expects current to be dominated by the injection of electron pairs
from the superconductor into the normal metal if the voltage
difference $V$ and the temperature $T$ are sufficiently low,
single-electron tunneling being forbidden by the energy required
to break a Cooper pair. Specifically, one expects two-electron
tunneling to dominate if $k_BT,eV \ll \Delta$, where $\Delta$ is
the zero-temperature superconductor gap. Simple and unquestionable
as this picture is, it is not clear how it can be quantitatively
described within the popular Bogoliubov - de Gennes (BdG)
quasiparticle scattering picture
\cite{blonder82,degennes66,beenakker95}. While it leaves the BCS
state unchanged, the emission of two electrons into the normal
metal involves the creation of two quasiparticles, something that
is not possible within the standard BdG formalism, where the
quasiparticle number is a good quantum number and the
quasiparticle scattering matrix is thus unitary. The conservation
of quasiparticle current is a consequence of the implicit
assumption contained in the conventional BdG scheme that the
reference chemical potential used to identify quasiparticles in
the normal metal is the superconductor chemical potential $\mu_S$.
However, as shown below, one does not need to be constrained by
such a choice.

In the mean field description of inhomogeneous superconductivity
provided by the BdG formalism, the Hamiltonian is given  by
\begin{equation}
\label{BdG-Hamiltonian} H=E_0 +
\sum_{n\sigma}\gamma^{\dagger}_{n\sigma} \gamma_{n\sigma},
\end{equation}
where $E_0$ is the condensate energy and
$\gamma^{\dagger}_{n\sigma}$ creates a quasiparticle of energy
$\varepsilon_n$, spin quantum number $\sigma$ and wave function
$[u_n({\bf r}),v_n({\bf r})]$ satisfying the BdG equations
\begin{eqnarray} \label{BdG-equations}
\left[ \begin{array}{cc}
H_0 -\mu_S & \Delta \\
\Delta^* & -H_0^* +\mu_S\end{array} \right]
\left[ \begin{array}{c} u_{n} \\
v_{n} \end{array} \right] = \varepsilon_{n}
\left[ \begin{array}{c} u_{n} \\
v_{n} \end{array} \right],
\end{eqnarray}
where $H_0=-\hbar^2\nabla^2/2m +U$ is the one-electron
Hamiltonian. In the standard convention one adopts solutions such
that $\varepsilon_n-\mu_S
> 0$. However, a fundamental property of the BdG equations
\cite{degennes66,sanc95,sanc97} is that, for every quasiparticle
$n,\sigma$ of energy $\varepsilon_n>\mu_S$, there exists another
solution $n',\sigma'$ with spin $\sigma'=-\sigma$, energy
$\varepsilon_{n'}=-\varepsilon_n+2\mu_S<\mu_S$ and wavefunction
$(u_{n'},v_{n'})=(-v_n^*, u_n^*)$. These two solutions are not
independent, since creating quasiparticle $n,\sigma$ is equivalent
to destroying quasiparticle $n',-\sigma$ \cite{sanc97}. More
specifically, $ \gamma^{\dagger}_{n\downarrow}
=\gamma_{n'\uparrow}$, and $\gamma_{n\uparrow}= -
\gamma^{\dagger}_{n'\downarrow}$.

In the case of a normal metal, where quasiparticles are pure
electron or pure holes, the above property implies that creation
of a quasiparticle of energy $\varepsilon_n>\mu_S$  and wave
function $(0,v_n)$ (i.e. a pure hole) corresponds to the
destruction of a quasiparticle of energy
$\varepsilon_n'=2\mu_S-\varepsilon_n<\mu_S$ and wave function
$(-v_n^*,0)$ (i.e. a pure electron). If $v_n({\bf r})\sim
\exp(i{\bf k}_h\cdot {\bf r})$, the existence of a hole of
momentum ${\bf k}_h$, with $k_h<k_F$, and energy
$\varepsilon>\mu_S$ corresponds to the absence of an electron in
the state of wave function $v_n^*({\bf r})\sim \exp(-i{\bf
k}_h\cdot {\bf r})$ with energy
$\varepsilon'=2\mu_S-\varepsilon<\mu_S$.

In a biased NS tunneling structure, the normal metal has a
different chemical potential $\mu_N=\mu_S-eV$. Without loss of
generality, we may assume $\mu_N<\mu_S$. If we release ourselves
from the standard BdG constraint of using $\mu_S$ as the reference
chemical potential even on the normal side, a clearer picture is
likely to emerge. We may decide that, in the energy range
$\mu_N<\varepsilon'<\mu_S$, we switch to the opposited convention
for the identification and labelling quasiparticles. In other
words, we decide to use $\mu_N$ as the reference chemical
potential. Translated to the example of the previous paragraph, we
pass to view the occupation of the hole-type quasiparticle state
of wave function $(0,v_n)$ and energy $\varepsilon>\mu_S$ as the
emptiness of the electron-type quasiparticle state of wave
function $(-v_n^*,0)$ and energy $\varepsilon'<\mu_S$. Conversely,
the absence of quasiparticles in $(0,v_n)$ is now viewed as the
occupation of $(-v_n^*,0)$, i.e. as the existence of an electron
with wave function $-v_n^*({\bf r})$ and energy $\varepsilon'$
between $\mu_N$ and $\mu_S$.

The consequences that this change of paradigm has on the way we
view transport through an NS interface can be more clearly
appreciated in Fig. \ref{2e-vs-Andreev}. In the standard BdG
picture represented in Fig. \ref{2e-vs-Andreev}(a), with $\mu_S$
as the reference chemical potential, the ``in" state is that of
many holes impinging on the NS interface from the N side, with
energies between $\mu_S$ and $\mu_S+eV$. Since ours is a tunneling
structure, normal reflection is the dominant scattering channel
and only one hole is Andreev reflected as an electron
(quasiparticle transmission is precluded at sufficiently low
temperatures and voltages). Thus the ``out" state is that of many
holes and only one electron moving away from the surface, all with
energies also between $\mu_S$ and $\mu_S+eV$, since quasiparticle
scattering is elastic. Given the unitary character of
quasiparticle scattering in the BdG formalism, the existence of an
outgoing electron requires the outgoing hole quasiparticle state
at the same energy to be empty, due to the incoming hole that
failed to be normally reflected. The absence of such an outgoing
hole is clearly shown in Fig. \ref{2e-vs-Andreev}(a).

If we now shift to $\mu_N$ as the reference chemical potential,
the picture is somewhat different. The many impinging holes on the
surfaces are now viewed as the absence of quasiparticles, i.e. the
``in" state is the vacuum of quasiparticles. The one electron that
emerged from a rare Andreev process continues to be viewed as an
occupied electron state, shown above $\mu_S$ in Fig.
\ref{2e-vs-Andreev}(b). The many outgoing holes of the BdG picture
are again viewed as an absence of quasiparticles. The {\it second
outgoing electron} that is needed to complete the picture of
two-electron emission corresponds to the empty outgoing hole state
of the BdG picture which originates from the {\it hole that failed
to be normally} (and thus specularly) {\it reflected}. It is shown
in Fig. \ref{2e-vs-Andreev}(b) with energy between $\mu_N$ and
$\mu_S$. As is known from the theory of quasiparticle Andreev
reflection, the outgoing electron of Fig. \ref{2e-vs-Andreev}(a)
follows the reverse path of the incident hole (conjugate
reflection). Therefore the two electrons in Fig.
\ref{2e-vs-Andreev}(b) have momenta with opposite parallel (to the
interface) components and the same perpendicular component, i.e.
they leave the superconductor forming a V centered around the axis
normal to the interface. Inclusion of the spin quantum number
completes the picture of two electrons emitted into the normal
metal in an entangled spin singlet state.

In summary, we have rigorously established the equivalence between
the pictures of Andreev reflection and two-electron emission,
noting that they emerge from {\it different choices of the
chemical potential} to which quasiparticles in the normal metal
are referred, $\mu_S$ in the standard BdG picture, and $\mu_N$ in
the scenario which contemplates the spontaneous emission of two
electrons. For simplicity, and because it better fits our present
need, we have focussed on the case of a tunneling structure.
However, the essence of our argument is of general validity. Here
we just note that, in the opposite case of a transmissive NS
interface \cite{blonder82,kummel69}, the same argument applies if,
exchanging roles, Andreev reflection passes to be the rule while
normal reflection becomes the exception. In that case, charge
accumulation and its accompanying potential drop, which are
generated by normal reflection \cite{sols99}, will be essentially
nonexistent.

Upon completion of this work, we have learned that the need to
change the normal metal vacuum to describe hole Andreev reflection
as electron emission has also been noted in Ref. \cite{samu03}.

\section{Formulation of the problem}

\begin{figure}
\includegraphics[width=8cm]{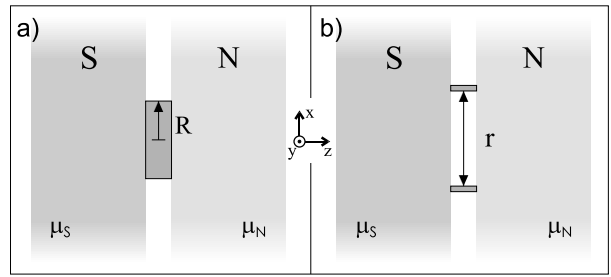}
\caption{Schematic lateral view of the NS tunneling structures
studied in this paper: (a) circular interface of arbitrary radius
$R$ and (b) interface made of two small holes at a distance $r$
from each other. The rest of the NS interface is assumed to be
opaque.} \label{SN-Structures}
\end{figure}

As has been said, an extensive body of literature has been written
on the various aspects of electron transport through a
normal-superconductor interface
\cite{andreev64,degennes66,griffin70,blonder82,byers95,torres99,
recher01,falci01,lesovik01,recher02,chtc02,feinberg03,recher03,lambert91,beenakker92,
kupka94,beenakker95,levy95,sanc95,tinkham96,kupka97,sanc97,sols99,
nakano94}. Generally those works have focussed on the case of
broad interfaces or point contacts \cite{beenakker95,levy95}. Our
goal here is to analyze the current of spin entangled Cooper pairs
from a BCS bulk superconductor into a bulk normal metal through an
arbitrarily shaped insulating junction in the tunnel limit. Apart
from the desire to explore novel types of NS structures, we are
also motivated by the need to investigate in depth the
two-electron emission picture, which is likely to be useful in the
design of quantum communication devices. We wish to consider
explicitly geometries of the sort depicted in Fig.
\ref{SN-Structures}, i.e. a 2D planar interface of arbitrary
radius $R$, presented in Fig. \ref{SN-Structures}(a), and two
small orifices separated by a distance $r$, shown in Fig.
\ref{SN-Structures}(b). It is assumed that, outside the designed
region, the interface is opaque to the flow of electrons. For
simplicity, both the normal and the superconducting electrodes are
taken to be ballistic. An advantage of the tunneling regime is
that the proximity effect may be neglected, i.e. we assume that
the gap function drops sharply at the NS interface and that
self-consistency in the gap may be safely neglected \cite{sanc97}.
Another benefit is that we deal at most with two chemical
potentials, since the low scale of tunneling currents guarantees
that the normal metal is close to equilibrium \cite{sols99} and
that no phase slips develop within the superconductor
\cite{sanc01}. Inelastic processes at the interface will also be
ignored \cite{kirtley92}.

We are interested in a conventional (s-wave) superconductor
because it may act as a natural source of spin-entangled
electrons, since its electrons form Cooper pairs with singlet spin
wave functions and may be injected into a normal metal. The
superconductor, which is held at a chemical potential $\mu_S$, is
weakly coupled by a tunnel barrier to a normal metal which is held
at $\mu_N$. By applying a bias voltage $V=(\mu_S-\mu_N)/e$ such
that $eV>0$, transport of entangled electrons occurs from the
superconductor to the normal metal. We focus on the regime
$k_BT\ll eV\ll\Delta$. Since $\delta \equiv \Delta/E_F \sim
10^{-4}$ in a conventional superconductor, rearrangement of the
potential barrier due to the voltage bias can be also neglected.
However, the effect of a finite, small $\delta$ will often be
tracked because pairing correlations (and thus nonlocal
entanglement) decays on the scale of the coherence length $\xi_0$,
which is finite to the extent that $\Delta$ is nonzero. For
convenience we assume that the superconductor normal-state
properties $(m,k_F,\rm{etc.})$ are the same as for an ordinary
metal.

We will use a tunneling Hamiltonian approach and explicitly
consider the emission of two electrons from the superconductor, a
viewpoint that will be mandatory in contexts where the late
evolution of correlated electron pairs in the normal metal is to
be investigated.

\section{Three-dimensional tunneling Hamiltonian}

The Bardeen model for electron tunneling \cite{bardeen61} assumes
that a system made up of two bulk metals connected through an
insulating oxide layer can be described by the Hamiltonian
\begin{equation}
H=H_L+H_R+H_T\,.
\end{equation}
Here $H_L$ and $H_R$ are the many-body Hamiltonians for the
decoupled (i.e. unperturbed) electrodes, the superconductor being
on the left and the normal metal on the right. The connection
between both electrodes is described by the tunneling term $H_T$
(see e.g. Ref. \cite{mahan00}):
\begin{equation}
\label{tunnel-Hamiltonian}
H_T=\sum_{\mathbf{k}\mathbf{q}\sigma}\,T_{\mathbf{k}\mathbf{q}}
\,c^{\dagger}_{\mathbf{k}\sigma}\,c_{\mathbf{q}\sigma}+\textrm{H.c.}
\end{equation}
Here $c^{\dagger}_{\mathbf{k}\sigma}$ is the creation operator in
the normal metal of the single-particle state of orbital quantum
number $\mathbf{k}$ and spin $\sigma$, whereas
$c_{\mathbf{q}\sigma}$ destroys state $\mathbf{q},\sigma$ in the
superconductor and $T_{\mathbf{k}\mathbf{q}}$ is the matrix
element connecting both states. We assume a perfect interface
defined by a square barrier $U({\bf
r},z)=U_0\Theta(z+w/2)\Theta(w/2-z)]$ (hereafter {\bf r} refers to
the in-plane coordinate).

If $\chi_{\mathbf{q}}(\mathbf{r},z)$ are the left-side stationary
waves for a potential step $U_L(\mathbf{r},z)=U_0\Theta(z+w/2)$
and $\chi_{\mathbf{k}}(\mathbf{r},z)$ behaves similarly for
$U_R(\mathbf{r},z)=U_0\Theta(w/2-z)$, Bardeen  \cite{bardeen61}
showed
\begin{widetext}
\begin{equation}
\label{Bardeen-hopping}T_{\mathbf{k}\mathbf{q}}=\frac{-\hbar^2}{2m}
\int d\mathbf{r}\,
\left[\chi^{\ast}_{\mathbf{k}}(\mathbf{r},z)\frac{\partial}{\partial
z}\chi_{\mathbf{q}}(\mathbf{r},z)-
\chi_{\mathbf{q}}(\mathbf{r},z)\frac{\partial}{\partial
z}\chi^{\ast}_{\mathbf{k}}(\mathbf{r},z)\right]_{z=z_{0}}\,,
\end{equation}
\end{widetext}
where $z_0$ lies inside the barrier, i.e.
$J_{\mathbf{k}\mathbf{q}}(\mathbf{r},z_0)\equiv(i/\hbar)T_{\mathbf{k}\mathbf{q}}$
is the matrix element of the $z$ component of the current density
operator. Due to charge conservation, $J_{\mathbf{k}\mathbf{q}}$
is independent of the choice of point $z_0\in[-w/2,w/2]$. The
unperturbed wave functions are of the form
\begin{equation} \label{wavefunction}
\chi_{\mathbf{k}}(\mathbf{r},z)=
\frac{e^{i\mathbf{k}_{\parallel}\mathbf{r}}}{\sqrt{A}}
\varphi_{k_z}(z),
\end{equation}
where the exact shape of $\varphi_{k_z}(z)$ depends on the barrier
height. Thus,
\begin{equation}
\label{hopping}
T_{\mathbf{k}\mathbf{q}}=\frac{\tau}{\sqrt{\Omega_{L}\Omega_{R}}N(0)}
\:\delta(\mathbf{k}_{\parallel}-\mathbf{q}_{\parallel})\,L(k_{z},q_{z})\,.
\end{equation}
Hereafter, the volume of each metal $\Omega_{L,R}$ is taken equal
to $\Omega=A L$, $A$ being the area of the interface and $L$ the
length of each semi-infinite metal. $N(0)$ is the 3D one-spin
electronic density of states of the normal metal at the Fermi
level: $N(0)=k_F^3/4\pi^2E_F$. We define the {\it transparency} of
the barrier as
\begin{equation}
\tau\equiv4\sqrt{\frac{E_{F}}{U_{0}}}e^{-p_0w}\,,
\end{equation}
where $p_0\equiv \sqrt{2mU_{0}}/\hbar$. In the particular case
$p_0w \gg 1$ and $E_F\ll U_0$, $\tau$ coincides with the
probability amplitude that an electron with perpendicular energy
$E_z=E_F$ traverses the barrier. $L(k_{z},q_{z})$ in Eq.
(\ref{hopping}) captures the dependence of the hopping energy on
the $z$ momentum component. Some authors take it as constant, but
we shall argue in section IX that its $k_z$, $q_z$ dependence is
crucial for a sound description of 3D transport problems.

For a square barrier, we may define $u\equiv U_0/E_{F}$,
$\varrho_z\equiv q_z/k_F$, $\kappa_z\equiv k_z/k_F$, and write
\begin{eqnarray}
\label{L}
L(k_{z},q_{z})=k_{z}\,q_{z}\,a(\kappa_z,\varrho_z)\,\exp{\{p_0w
\left[1-a(\kappa_z,\varrho_z)\right]\}}\,,
\end{eqnarray}
where
\begin{eqnarray}
\label{a}a(x,y)&\equiv&[b(x)+b(y)]/2\,,\\
\label{b}b(x)&\equiv&\sqrt{1-x^2/u}\,.
\end{eqnarray}
For high barriers ($u\rightarrow\infty$) we have
$a(x)\rightarrow1$. Then,
\begin{equation}
\label{hopping-u-inf}T_{\mathbf{k}\mathbf{q}} \simeq
\frac{\tau}{\Omega
N(0)}\,\delta(\mathbf{k}_{\parallel}-\mathbf{q}_{\parallel})\,k_{z}\,q_{z}\,.
\end{equation}
If we make $U_0\rightarrow \infty$ while keeping the electron
transmission probability finite, we are implicitly assuming that
the barrier becomes arbitrarily thin ($w\rightarrow 0$), i.e. we
are taking it to be of the form $V(z)=H\delta(z)$, as popularized
in Ref. \cite{blonder82}. On the other hand, since the height of
the barrier is judged in relation to the perpendicular energy
$E_z\leq E_F<U_0$, it is clear that, given $U_0$ and $w$, Eq.
(\ref{hopping-u-inf}) becomes correct for sufficiently small
$k_z,q_z$. In other words, $T_{\mathbf{k}\mathbf{q}}$ behaves
identically for $u\rightarrow\infty$ or $k_z,q_z\rightarrow 0$. As
a consequence, such bilinear dependence of
$T_{\mathbf{k}\mathbf{q}}$ for sufficiently small $k_z,q_z$  may
be expected to hold for arbitrary barrier profiles within the
tunneling regime. We note that Eq. (\ref{hopping-u-inf}) differs
from the result obtained in Ref. \cite{gray65} for the low energy
hopping.

\subsection{Validity of the tunneling Hamiltonian model: momentum cutoff}

We wish to quantify the idea that a perturbative treatment of
Bardeen's tunneling Hamiltonian is valid only when it involves
matrix elements between weakly coupled states
\cite{bardeen61,prange63}.

The transmission probability for a low energy electron incident
from the left can be written
\begin{equation}
T(E_z) = W_{\mathbf{q}}/J_{\mathbf{q}}\,,
\end{equation}
where $J_{\mathbf{q}}$ is the current density carried by the
incoming component of the stationary wave {\bf q}, and
\begin{equation}
W_{\mathbf{q}}=
\frac{2\pi}{\hbar}\sum_\mathbf{k}
|T_{\mathbf{k}\mathbf{q}}|^2\delta(E_{\mathbf{k}}-E_{\mathbf{q}})\,
\end{equation}
is the tunneling rate. Using Eqs. (\ref{hopping}) and (\ref{L}),
Bardeen's theory yields
\begin{equation} \label{tunneling-transmission}
T(E_z)=16\frac{E_z}{U_0}\left(1-\frac{E_z}{U_0}\right)e^{-2p_zw}\,,
\end{equation}
where $p_z=\sqrt{2m(U_0-E_z)}/\hbar$. On the other hand, an exact
calculation \cite{galindo90} recovers the tunneling result
(\ref{tunneling-transmission}) for $E_z<U_0$
\cite{com-duke,duke69} if we make the approximation
\begin{equation}
\label{tunnel-condition}\sinh(p_zw)\approx\cosh(p_zw)\approx
e^{p_zw}/2\,.
\end{equation}
Thus we adopt as a criterion of the validity of Bardeen's
approximation that Eq. (\ref{tunnel-condition}) holds, which from
(\ref{tunneling-transmission}), implies $T(E_F)\ll 1$. This
defines an upper energy cutoff $E_c$ in the various sums over
electron states, which is the maximum energy for which the
approximation (\ref{tunnel-condition}) is valid. For the square
barrier, $E_c \simeq U_0 -\hbar^2/2mw^2$.

For processes described by amplitudes which are first order in
$H_T$, and as long as $U_0$ is high enough compared to $E_F$ to
fulfill condition (\ref{tunnel-condition}) for all relevant $E_z$,
all electron momenta lie within the applicability of the tunnel
limit and we may use the tunneling Hamiltonian safely. That is the
case of the tunnel current through a NN interface or the
quasiparticle tunnel current through a NS interface.

The situation is different for transport through a NS interface,
since it requires the coherent tunneling of two electrons. Then,
the leading contribution to the tunneling amplitude is quadratic
in $H_T$ and the final transmission probability is sensitive to
the existence of intermediate virtual states where only one
electron has tunneled and a quasiparticle above the gap has been
created in the superconductor. Unlike the weighting factors of the
initial and final states, which are controlled by the Fermi
distribution function, the contribution of the virtual
intermediate states decays slowly with energy and the cutoff $E_c$
may be reached. In section V we show that there are two cases
where the cutoff can be safely neglected, namely, the limit of
high barrier ($u \gg 1$) and the limit of small gap ($\delta \ll
1$).

\subsection{Tunneling Hamiltonian in real space}

One of our main goals is to investigate transport through
tunneling interfaces of arbitrary shape \cite{chen90} that are
otherwise uniform. For that purpose we need a reliable tunneling
Hamiltonian expressed in real space. Our strategy will be to
rewrite Eq. (\ref{tunnel-Hamiltonian}) as an integral over the
infinite interface and postulate that a similar Hamiltonian, this
time with the integral restricted to the desired region, applies
to tunneling through the finite-size interface. The discontinuity
between the weakly transparent interface and the completely opaque
region causes some additional scattering in the electronic wave
functions that enter the exact matrix element. However, this
effect should be negligible in the tunneling limit. In fact, we
provide in Appendix A  an independent derivation of the continuum
results shown in this section which starts from a discrete
tight-binding Hamiltonian.

Thus, in (\ref{tunnel-Hamiltonian}) we introduce the
transformations
\begin{eqnarray}
c^{\dagger}_{ \mathbf{k}\sigma}=\int_{A} d\mathbf{r}
\int\!\!dz\;\chi_{\mathbf{k}}(\mathbf{r},z)
\:\psi^{\dagger}_{N}(\mathbf{r},z;\sigma)\\
c_{\mathbf{q}\sigma}=\int_{A} d\mathbf{r} \int dz \;
\chi^*_{\mathbf{q}}(\mathbf{r},z)\:\psi_{S}(\mathbf{r},z;\sigma)
\end{eqnarray}
where the wave functions $\chi_{\mathbf{q}}$ and
$\chi_{\mathbf{k}}$ are, respectively, solutions of $H_L$ and
$H_R$ and are given in (\ref{wavefunction}).
$\psi^{\dagger}_{N}(\mathbf{r},z;\sigma)$ and
$\psi_{S}(\mathbf{r},z;\sigma)$ are the field operator in the
normal and superconducting metals.

Invoking Eq. (\ref{hopping}) and the completeness of plane waves
in the $x,y$ plane [which yields a term
$\delta(\mathbf{r}-\mathbf{r}')$], we obtain
\begin{widetext}
\begin{equation} \label{ham-real-space-1}
H_{T}=\sum_{\sigma}\frac{\tau}{4 \pi^{2}N(0)}\int_{A}
d\mathbf{r}\int dz'\int dz\:\widetilde{L}(z,z')\:
\psi^{\dagger}_{N}(\mathbf{r},z;\sigma)\,\psi_{S}(\mathbf{r},z';\sigma)+\textrm{H.c.}\,,
\end{equation}
\end{widetext}
where
\begin{equation}
\label{L-tilda}\widetilde{L}(z,z')=\frac{1}{L_{z}}
{\sum_{k_{z},q_{z}>0}}\varphi_{k_z}\!(z)\,\varphi^{*}_{q_z}\!(z')\,L(k_z,q_z)\,.
\end{equation}

Since the initial Hamiltonian (\ref{tunnel-Hamiltonian}) connects
states which overlap in a finite region below and near the
barrier, it is logical that the real space Hamiltonian
(\ref{ham-real-space-1}) is non-local in the $z$-coordinate. An
interesting limit is that of a high and (to keep transmission
finite) thin barrier, i.e. the delta barrier limit. Then, the
perpendicular wave functions can be precisely written
\begin{equation} \label{perpendicular-wave-function}
\varphi_{k_z}\!(z)=\sqrt{2/L}\sin(k_zz), \;\;z\ge 0
\end{equation}
and similarly for the left electrode. We introduce such wave
functions in Eq. (\ref{L-tilda}) and invoke the identity
(hereafter $L\rightarrow\infty$)
\begin{equation} \label{delta-derivative}
\frac{1}{L}\sum_{k_z>0} k_z\sin(k_z z)= -\delta'(z),
\end{equation}
where the volume per orbital in $k_z$-space is $\pi/L$. Then, to
leading order in $u^{-1}\ll 1$, Eq. (\ref{ham-real-space-1})
yields
\begin{widetext}
\begin{eqnarray}
\label{Ht-realspace-u-inf}
H_{T}=\sum_{\sigma}\frac{\tau}{8\pi^{2}N(0)}\int_A\!
d\mathbf{r}\,\left.\frac{\partial
\psi^{\dagger}_{N}(\mathbf{r},z;\sigma)}{\partial
z}\right|_{z\rightarrow 0^+} \left.\frac{\partial
\psi_{S}(\mathbf{r},z';\sigma)}{\partial z'}\right|_{z'\rightarrow
0^-}+\textrm{H.c.}
\end{eqnarray}
\end{widetext}

If we replace the thermodynamic area $A$ by a specific finite
area, the real space Hamiltonian (\ref{Ht-realspace-u-inf}) can be
used to describe tunneling through interfaces of arbitrary shape.
As we have said, in Appendix A we provide an alternative
derivation which makes Eq. (\ref{Ht-realspace-u-inf}) appear as
the natural continuum limit of the hopping Hamiltonian in a
regularized tight-binding representation. We note that the
tunneling Hamiltonian (\ref{Ht-realspace-u-inf}) may also be
obtained if the r.h.s. of Eq. (\ref{perpendicular-wave-function})
is replaced by a plane wave representation.

>From Eq. (\ref{Ht-realspace-u-inf}) we conclude that apparently
reasonable choices of local tunneling Hamiltonian such as those
$\propto \int\psi_S^{\dagger}\psi_N$ lead to unphysical results in
3D. This point will be discussed in depth in section IX.

To describe tunneling in real space, rather than starting from
Hamiltonian (\ref{ham-real-space-1}), or its limiting version
(\ref{Ht-realspace-u-inf}), it is  more convenient in practice to
go back to Eq. (\ref{hopping}) and make the replacement
$\delta(\mathbf{k}_{\parallel}-\mathbf{q}_{\parallel})\rightarrow(2\pi)^{-2}\int_Ad\mathbf{r}\,
e^{i(\mathbf{k}_\parallel-\mathbf{q}_\parallel)\cdot \mathbf{r}}$,
with $A$ finite. Then Eqs. (\ref{ham-real-space-1}) and
(\ref{L-tilda}) may equivalently be written
\begin{equation} \label{practical-Hamiltonian}
H_T\!=\!\sum_{\mathbf{k}\mathbf{q}\sigma}\frac{\tau(2\pi)^{-2}}{N(0)\Omega}\int_A\!
d\mathbf{r}\,e^{i(\mathbf{k}_\parallel-\mathbf{q}_\parallel)\cdot\mathbf{r}}L(k_z,q_z)
c^{\dagger}_{\mathbf{k}\sigma}\,c_{\mathbf{q}\sigma}+\textrm{H.c.}
\end{equation}
If we make $L(k_z,q_z)=k_zq_z$, it is easy to prove that
(\ref{practical-Hamiltonian})  becomes (\ref{Ht-realspace-u-inf}).

\section{Perturbative calculation of the stationary current}

Following Ref. \cite{recher01}, we write the stationary electron
current from the superconductor to the normal metal as
\begin{equation}
\label{current} I_{\rm NS}=2e\sum\limits_{if}W_{fi}\,\rho_{i}\, ,
\end{equation}
where $W_{fi}$  is the transition rate at which {\it two}
electrons tunnel from the superconductor into the normal metal,
and $\rho_i$ is the stationary distribution accounting for a
chemical potential difference $eV$ between the two electrodes. We
calculate the transition rate with a $T$-matrix approach
\cite{merzbacher98},
\begin{equation}
\label{Wfi}W_{fi}=\frac{2\pi}{\hbar}\,|\langle
f|\,\hat{T}(\varepsilon_i)\,|i\rangle|^2\,\delta(\varepsilon_f-\varepsilon_i)\,.
\end{equation}
The $T$-matrix can be written as a power series in the tunnel
Hamiltonian $H_{T}$,
\begin{equation}
\label{series} \hat{T}(\varepsilon)=H_{T}+H_{T}\sum_{n=1}^{\infty}
[G_{0}(\varepsilon)H_{T}]^n \,
\end{equation}
where $G_{0}(\varepsilon)=(\varepsilon-H_{0}+i0^+)^{-1}$ is the
retarded Green function for the decoupled system.

At zero temperature the initial state is
$|i\rangle=|\textrm{F}\rangle\otimes|\textrm{BCS}\rangle$, where
$|\textrm{F}\rangle$ is the filled Fermi sea ground state of the
normal metal and $|\textrm{BCS}\rangle$ is the BCS ground state of
the superconductor. The state $|i\rangle$ is the vacuum of
quasiparticles if these are referred to $\mu_S$ in the
superconductor and to $\mu_N=\mu_S-eV$ in the normal metal (see
section II). In the final state
\begin{equation} \label{singlet}
|f\rangle=\frac{1}{\sqrt{2}}
(c^{\dagger}_{\mathbf{k}_{1}\uparrow}c^{\dagger}_{\mathbf{k}_{2}\downarrow}-
c^{\dagger}_{\mathbf{k}_{1}\downarrow}c^{\dagger}_{\mathbf{k}_{2}\uparrow})|i\rangle\,,
\end{equation}
i.e. the superconductor remains unperturbed within the BCS
description, since an entire Cooper pair has been removed, and two
singlet-correlated electrons hover above the normal Fermi sea
\cite{comm-Fermi}. In the ensuing discussion we take
$\varepsilon_i=2\mu_S\equiv 0$.

Since we wish to focus on the regime $k_BT\ll eV\ll\Delta$, single
electron emission is forbidden due to energy conservation, because
it requires the breaking of a Cooper pair. Therefore, to leading
order in $H_T$, we may approximate
\begin{equation} \label{T0}
\hat{T}(0)\approx T^{''}\!\equiv H_{T}\,G_{0}\,H_{T}\,
\end{equation}
and so we write
\begin{equation}
\label{amplitude} \langle
f|\hat{T}(0)|i\rangle=\frac{1}{\sqrt{2}}\,\langle
(c_{\mathbf{k}_{2}\downarrow}c_{\mathbf{k}_{1}\uparrow}-
c_{\mathbf{k}_{2}\uparrow}c_{\mathbf{k}_{1}\downarrow})\,T^{''}
\rangle\, .
\end{equation}

We insert a complete set of single-quasiparticle (virtual) states,
i.e. $1\!\!1=\sum_{\mathbf{k}\mathbf{q}\sigma\sigma'}
\gamma^{\dagger}_{\mathbf{q}\sigma}c^{\dagger}_{\mathbf{k}\sigma'}|i\rangle\langle
i|c_{\mathbf{k}\sigma'}\gamma_{\mathbf{q}\sigma}$, between the two
$H_{T}$ in (\ref{T0}) and we use the fact that the resulting
energy denominator $|i0^+-\xi_{\mathbf{k}}-E_{\mathbf{q}}|\approx
|E_{\mathbf{q}}|$, since
$\xi_{\mathbf{k}}\equiv\hbar^2k^2/2m-\mu_N\rightarrow 0$ when $e
V\rightarrow 0$. To see this, one must note that the energy
conservation implies $\varepsilon_{f}=\varepsilon_{i}$; therefore,
$\varepsilon_{f}=\xi_{\mathbf{k}_{1}}+\xi_{\mathbf{k}_{2}}-2 e
V=2\mu_{S}=0$. Thus, when $e V\rightarrow 0$, one may write
$\xi_{\mathbf{k}_{1}}\approx -\xi_{\mathbf{k}_{2}}\approx0$ . We
also make use of
$u_{\mathbf{q}}v_{\mathbf{q}}=u_{-\mathbf{q}}v_{-\mathbf{q}}$.
Finally, we get
\begin{eqnarray}
&&\label{Tfi} \langle
f|\hat{T}(0)|i\rangle=\nonumber\\
&&~~=2\sqrt{2}\langle
c_{\mathbf{k}_2\uparrow}c^{\dagger}_{\mathbf{k}_2\uparrow}
c_{\mathbf{k}_1\downarrow}c^{\dagger}_{\mathbf{k}_1\downarrow}\rangle
{\sum_{\mathbf{q}}}\frac{\langle
c_{\mathbf{q}\uparrow}c_{\mathbf{-q}\downarrow}\rangle}{E_{\mathbf{q}}}
T_{\mathbf{k}_{1}\mathbf{q}} T_{\mathbf{k}_{2},\mathbf{-q}}~~~~~~
\end{eqnarray}
where
$E_{\mathbf{q}}=[(\hbar^2/2m)^2(q^2-k_F^2)^2+\Delta^2]^{1/2}$ is
the quasiparticle energy and $F_\mathbf{q}\equiv\langle
c_{\mathbf{q}\uparrow}c_{\mathbf{-q}\downarrow}\rangle$ is the
condensation amplitude in the state $\mathbf{q}$
\cite{degennes66}.

At zero temperature we have $F_\mathbf{q} =\Delta/2
E_{\mathbf{q}}$. Thus, in the summation of Eq. (\ref{Tfi}), the
contribution of high energy virtual states is weighted by the
Lorentzian $F_\mathbf{q}/E_\mathbf{q}$, of width  $\Delta$ and
centered around $E_F$. We already mentioned in the previous
section the need for a high-energy cutoff $E_c$ to prevent the
inclusion of states for which the tunneling approximation is not
valid. However, in the limit $\Delta/E_F \rightarrow 0$, the
Lorentzian distribution becomes a delta function and the results
are independent of the cutoff, which can be safely taken to
infinity. A similar situation is found in the limit $U_0 \gg E_F$,
for which the sum in Eq. (\ref{Tfi}) converges before reaching the
energy $E_c$ above which Bardeen's approximation is no longer
valid. In any of these two limits ($\Delta/E_F,E_F/U_0 \ll 1)$, it
is correct to take $E_c \rightarrow \infty$.

\section{Total current and angular distribution through a broad interface}

\begin{figure}
\includegraphics[width=5cm]{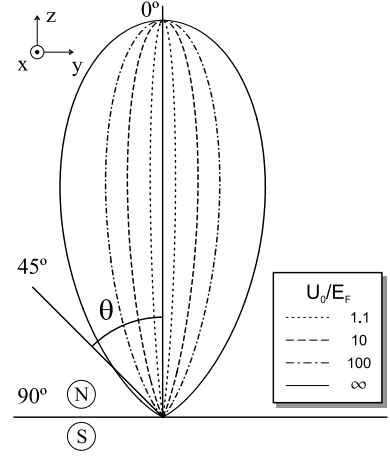}
\caption{Angular dependence of the normalized tunnel current
ranging from $U_{0}/E_{F}\rightarrow\infty$ for the outer
($\cos^5\theta$) curve to $U_{0}/E_{F}=1.1$ for the inner one.
Finite barriers have a width $w=5\lambda_F$. Observe how the
angular distribution focalizes around the perpendicular direction
as the barrier hight decreases.} \label{Dep-Angular}
\end{figure}

The current through a NS junction is most easily calculated when
the interface section is much bigger than $\lambda_F$. We shall
refer to it as the broad interface or thermodynamic limit. Its
detailed understanding is of interest for later reference in the
investigation of finite size interfaces. For $k_BT\ll eV\ll
\Delta$, the set of equations in the previous section yields
\begin{equation}
\label{tunnel-current} I_{\rm{NS}}=
I_{V}\frac{\tau^{4}}{2}\int_{0}^{\pi/2}\!d\theta\,\sin\!\theta\,
g(\theta) \,,
\end{equation}
where $g(\theta)$ is the angular distribution ($\theta$ being the
angle between the outgoing electron momentum and the direction
normal to the interface) and
\begin{equation}
\label{IV} I_{V}\equiv\ \frac{1}{2}\,e^{2}\,V N(0)\,v_{F}\,A\, =
J_V A \ ,
\end{equation}
with $V$ the applied voltage, $A$ the interface area, and $v_{F}$
the Fermi velocity.  Eq. (\ref{IV}) may be written as
$I_V=(2e^2/h)NV$, where $N=Ak_F^2/4\pi$ is the number of
transverse channels that fit in an interface of area $A$. Thus,
$I_V$ can be interpreted as the current that would flow through a
transmissive [$T(E_z)=1$ for all $E_z$] normal-normal interface
with the same area and subject to the same voltage bias. The
$\tau^4$ dependence of $I_{\rm NS}$ reflects the simultaneous
tunneling of two electrons.

Using the previous definition $\varrho_z\equiv q_z/k_F$, the
angular distribution for the current through an arbitrary square
barrier is \cite{comm-morten99,mortensen99}
\begin{widetext}
\begin{eqnarray} \label{g-theta-general}
g(\theta)=2\cos^3\theta \,
e^{2p_0w[1-b(\cos\theta)]}\left[\frac{2}{\pi}\int_0^{\varrho_c}
\!\!\!d\varrho_z\:\frac{\delta}{(\varrho_z^2-\cos^2\theta)^2+\delta^2}
\:\varrho_z^2\:\left[a(\varrho_z,\cos\theta)\right]^2
\:e^{p_0w[1-b(\varrho_z)]}\right]^2\,,\nonumber\\
\end{eqnarray}
\end{widetext}
where $\varrho_c=\sqrt{E_c/E_F}$ is the cutoff and the functions
$a$ and $b$ were defined in Eqs. (\ref{a}) and (\ref{b}).

For $\delta\rightarrow0$, we have
\begin{equation}
\lim_{\delta\rightarrow0}\int_{0}^{\varrho_c}\!\!\!d\varrho\,\frac{\delta}{(\varrho^2\!-\!x^2)^2\!+\delta^2}\;\varrho
\,f(\varrho)=\frac{\pi}{2}\,f(x)\, ,
\end{equation}
if, as is the case, $\varrho_c > 1 \geq |x|$. Therefore, in the
limit $\delta\rightarrow0$, Eq. (\ref{g-theta-general}) yields
\begin{equation}
\label{angular-dep}g(\theta)= 2\,
e^{4p_0w[1-b(\cos\!\theta)]}[b(\cos\!\theta)]^{4}
\cos^{5}\!\theta\,.
\end{equation}

For large barriers ($u \gg 1$) and finite $\delta$ we find (with
$\varrho_c \rightarrow \infty$)
\begin{eqnarray} \label{g-high-u}
g(\theta)&=&
\cos^{5}\!\theta+\cos^{3}\!\theta\sqrt{\cos^{4}\!\theta+\delta^{2}}\\
\label{g-cos-5-theta}&=&2\cos^{5}\theta\left[1+\mathcal{O}(\delta^2)\right].
\end{eqnarray}
Combining Eqs. (\ref{tunnel-current}) and (\ref{g-high-u}), we
obtain for the total current
\begin{eqnarray}
\label{ISNu-inf}I_{\rm{NS}}&=&
\frac{1}{12}I_{V}\tau^4\left[1+(1+\delta^2)^{3/2}-\delta^3\right]
\\ \label{ISNu-inf-small-delta}
&=&\frac{1}{6}I_{V}\tau^4\left[1+\mathcal{O}(\delta^2)\right].
\end{eqnarray}

However, if the cutoff $\varrho_c$ remains finite, Eq.
(\ref{ISNu-inf-small-delta}) must be replaced by
\begin{equation}
\label{ISNu-inf-cut-finite} I_{\rm{NS}}=
\frac{1}{6}I_{V}\tau^4\left[1-\frac{24}{5\pi}\frac{\delta}{\varrho_c}+
\mathcal{O}(\delta^2)\right],
\end{equation}
i.e. a finite cutoff qualitatively affects the leading
low-$\delta$ dependence of $I_{\rm NS}$.

The underlying physics goes as follows. The product of hopping
matrix elements appearing in (\ref{Tfi}) satisfies
\begin{equation}
T^{\ast}_{\mathbf{k_{1}}\mathbf{q}}T^{\ast}_{\mathbf{k_{2}},\mathbf{-q}}\propto
k_{1z}\:k_{2z}\:q_z^2\:\delta(\mathbf{k}_{1\parallel}+\mathbf{k}_{2\parallel})\delta(\mathbf{k}_{1\parallel}-\mathbf{q}_{\parallel})\,.
\end{equation}
Thus, when crossing the barrier, electrons forming a Cooper pair
of momenta ($\mathbf{q},-\mathbf{q}$) undergo the following
process: Their opposite interface-parallel momenta are conserved
($\mathbf{k}_{1\parallel}=\mathbf{q}_{\parallel}$ and
$\mathbf{k}_{2\parallel}=-\mathbf{q}_{\parallel}$). By contrast,
one of their perpendicular momentum components (more specifically,
the negative one pointing away from the interface) is reversed so
that both electrons enter the normal metal with perpendicular
momenta $k_{1z},k_{2z}>0$. In the limit of $eV\rightarrow0$ the
modulus difference between $k_{1z}$ and $k_{1z}$ is negligible.
This means that the electron current through a broad interface
will propagate into the normal lead in the form of two rays which
are symmetric with respect to the direction normal to the
interface. Due to axial symmetry, $g$ is only a function of the
zenithal angle $\theta\in[0,\pi/2]$.

The normalized angular distributions for several barrier heights
are depicted in Fig. \ref{Dep-Angular} in the limit
$\delta\rightarrow0$. The lowest barrier which we have considered
has $u=1.1$. This means that, for a typical value of $E_F=5$ eV,
the difference between the height of the barrier and the Fermi
energy is $0.5$ eV, i.e. large enough to ensure that the junction
operates in the tunneling regime. In Fig. \ref{Dep-Angular},
finite-height barriers are taken to have a width $w=5\lambda_F$.
For large $U_0$ we reproduce the analytical $\cos^{5}\theta$
behavior given in Eq. (\ref{g-cos-5-theta}). As the barrier height
decreases, the angular distribution becomes more focussed in the
forward direction because transmission is more sensitive to the
perpendicular energy. Thus the relative fraction of Fermi surface
electrons crossing the interface with $E_z$ close to the highest
value $E_F$ increases. That majority of transmitted electrons have
low parallel momenta and, accordingly, a characteristic parallel
wave length much larger than $\lambda_F$. We will see later that
this perpendicular energy selection bears consequences on the
length scale characterizing the dependence of the total current on
the radius of the interface.

In general, knowledge of the current angular distribution is
physically relevant, as one is ultimately interested in
directionally separating the pair of entangled electron beams for
eventual quantum information processing. To acquire a more
complete picture, we may compare the previous results with the
case of a NN interface. In that case the total tunnel current is
\begin{equation} \label{tunnel-NN-current}
I_{\rm{NN}}=
I_{V}\tau^{2}\int_{0}^{\pi/2}\!d\theta\,\sin\!\theta\,
g(\theta)\,,
\end{equation}
where $I_V$ is given in Eq. (\ref{IV}) and, for large $u$,
\begin{equation} \label{g-cos-3-theta}
g(\theta)=2\cos^3\theta\,.
\end{equation}
Thus we see that electron transport through a tunneling NN
interface also exhibits focussing which is however less sharp than
in the NS case [see Eq. (\ref{g-cos-5-theta})]. The term $\tau^2$
in Eq. (\ref{g-cos-3-theta}) reflects the dominance of
single-electron tunneling at the NN interface. Finally, we may
compare Eqs. (\ref{g-cos-5-theta}) and (\ref{g-cos-3-theta}) with
the $\cos \theta$ distribution law exhibited by electron current
in the bulk of a disordered wire \cite{comm1}.

\subsection{Connection with the multi-mode picture}

We could have derived the angular distributions in Eqs.
(\ref{tunnel-current}), (\ref{g-cos-5-theta}),
(\ref{tunnel-NN-current}) and (\ref{g-cos-3-theta}) following the
scattering theory of conduction in normal \cite{landauer57} and
normal-superconducting \cite{beenakker92} multichannel wires. For
an NN interface we can write the relation between conductance and
transmission probabilities at the Fermi energy as
\begin{equation}
\label{GN}G_{\rm
NN}=\frac{2e^2}{h}\sum_{n=1}^{N}A_n=\frac{2e^2}{h}\sum_{n=1}^{N}T_n\,,
\end{equation}
where $A_n$ $(n=1,2,... N)$ are the eigenvalues of the
transmission matrix $tt^+$ at the Fermi energy and
$T_n\equiv\sum_{m=1}^N|t_{nm}|^2$ are the modal transmission
probabilities at the same energy \cite{beenakker95}, which is what
we calculate. The exchangeability of $\sum_nA_n$ and $\sum_nT_n$
reflects the invariance of the trace \cite{comm-trace}. Now
consider the transmission probability through a square barrier
given in Eq. (\ref{tunneling-transmission}). We replace
$T_n\rightarrow T(E_z)$. For $E_z/U_0 \ll 1$, we have
$T(E_z)\propto E_z\propto \cos^2\theta$. Moreover, the sum over
transverse modes can be replaced by an integration over the
zenithal angle, $\sum_n \rightarrow {\rm
cnst.}\times\int_0^{\pi/2} d\theta\sin\theta\cos\theta$.
Altogether, the angular distribution follows the $\cos^3\theta$
law expressed in Eq. (\ref{g-cos-3-theta}).

A similar line of argument can be followed for the Andreev current
through a NS interface, whose conductance is given by
\begin{equation}
\label{GNS}G_{\rm NS}=\frac{2e^2}{h}\sum_{n=1}^{N}
\frac{2A_n^2}{(2-A_n)^2}\,.
\end{equation}
As noted in Ref. \cite{beenakker95}, the equivalence invoked in
Eq. (\ref{GN}) is no longer applicable in Eq. (\ref{GNS}) because
of its nonlinearity. Nevertheless, in the tunneling limit one has
$A_n \ll 1$ and $G_{\rm NS}$ can thus be approximated as
\begin{equation}
\label{GNS-tunnel}G_{\rm
NS}\simeq\frac{2e^2}{h}\sum_{n=1}^{N}\frac{A_n^2}{2}
=\frac{2e^2}{h}\sum_{n=1}^N\frac{T_n^2}{2}\, ,
\end{equation}
where the second equality is possible because $\sum_nA_n^2=
{\rm{Tr}}(tt^+)^2$. Arguing as we did for the NN conductance, it
follows that $g(\theta) \propto \cos^5\theta$, which confirms Eq.
(\ref{g-cos-5-theta}). We note here that, in Refs.
\cite{beenakker92,beenakker95}, the {\it Andreev approximation}
was made whereby all the momenta involved are assumed to be equal
to $k_F$. In our language, this corresponds to taking
$\delta\rightarrow0$ in Eq. (\ref{g-theta-general}) and
thereafter.

Finally, comparison of Eqs. (\ref{GN}) and (\ref{GNS-tunnel}) also
illuminates the contrast between the factor $\tau^4/2$ in Eq.
(\ref{tunnel-current}) and the factor $\tau^2$ in Eq.
(\ref{tunnel-NN-current}).

\subsection{Universal relation between NN and NS tunneling conductances}

In the case of a normal interface with high barrier, the total
current can be integrated to yield
\begin{equation} \label{I-NN}
I_{\rm{NN}}=\frac{I_{V}\tau^{2}}{2}=\left(\frac{2e^2}{h}\right)\frac{\tau^{2}}{2}
NV\,.
\end{equation}
Thus $\tau^2/2$ is the average transmission per channel
\cite{comm-BTK,sharvin65}. In one dimension ($N=1$) one has
$I_{\rm NN}=(2e^2/h)V\tau^{2}$. Eqs. (\ref{IV}), (\ref{ISNu-inf}),
and (\ref{I-NN}) suggest the universal relation
\begin{equation} \label{universal}
\frac{I_V I_{\rm NS}}{(I_{\rm NN})^2}=\frac{G_V G_{\rm
NS}}{(G_{\rm NN})^2}=\frac{2}{3}\,,
\end{equation}
where $G_i=I_i/V$ ($i={\rm V,NS,NN}$). Eq. (\ref{universal})
indicates that knowledge of $G_{\rm NS}$ and $G_{\rm NN}$ may
allow us to infer $G_V$ and, from (\ref{IV}), the {\it effective
area} of a tunneling interface.

\subsection{Comparison with the quasiparticle scattering method}

Blonder et al. {\cite{blonder82} studied transport through a
one-dimensional NS interface modelled by a delta-barrier
one-electron potential [$U(z)=H\delta (z)$] by solving for the
quasiparticle scattering amplitudes. If the dimensionless
parameter $Z=mH/\hbar^2k_F$ is employed to characterize the
scattering strength of the barrier, the tunneling limit
corresponds to $Z\gg1$, for which they obtained $I_{\rm NS}^{\rm
1D}=I_V/Z^4$ assuming $k_BT\ll eV\ll\Delta\ll E_F$ (i.e. a
low-transmission regime in which Andreev reflection is the only
charge-transmitting channel). Later, Kupka generalized the work of
Ref. {\cite{blonder82} to investigate the sensitivity of Andreev
and normal reflection to the thickness of the barrier
\cite{kupka94} and to the presence of a realistic 3D geometry
\cite{kupka97}. For the case of a broad interface in the tunneling
limit he obtained $I_{\rm NS}^{\rm 3D}=I_V/6Z^4$. Therefore, Kupka
found a result identical to Eq. (\ref{ISNu-inf-small-delta}) (to
zeroth order in $\delta$) with $\tau$ replaced by $1/Z$. In fact,
it is easy to see that, in the case of a delta-barrier with
$Z\gg1$, the transparency defined in section IV is precisely
$\tau=1/Z$. Therefore, comparison of Eqs. (\ref{g-cos-5-theta})
and (\ref{ISNu-inf-small-delta}) with the results of Ref.
\cite{kupka97} completes the discussion of section II by
establishing the {\it quantitative equivalence} between the
pictures of quasiparticle Andreev reflection and two-electron (or
two-hole) emission. We note that, in Refs.
\cite{blonder82,kupka94,kupka97}, the Andreev approximation
($\delta \rightarrow 0$) was made.

\section{Current through a circular interface of arbitrary radius}

In this section we investigate transport through a circular NS
tunneling interface of arbitrary radius. The setup is as depicted
in Fig. \ref{SN-Structures}(a). To make the discussion more
fluent, lengthy mathematical expressions have been transferred to
Appendix B, leaving here the presentation of the main results,
which include some analytical expressions for the limit of small
gap and high barrier.

\subsection{Total current}

\begin{figure}
\includegraphics[width=8.5cm]{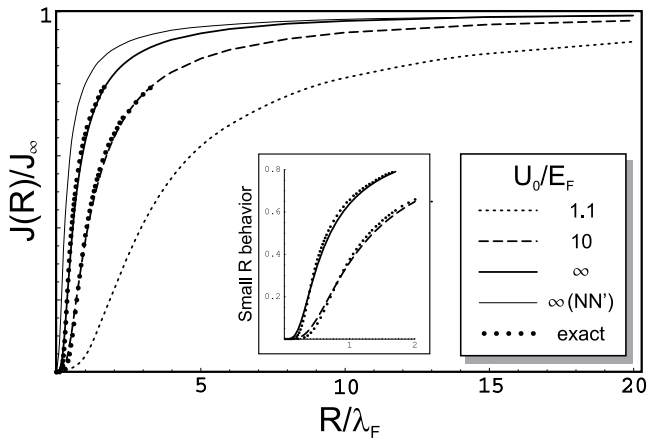}
\caption{Radial dependence of the normalized NS Andreev tunnel
current through a circular interface of radius $R$ for different
barrier heights. $J(R)\equiv I(R)/\pi R^2$ and $J_{\infty}$ is the
current density in the thermodynamic limit. Finite barriers have a
width $w=5\lambda_F$. Everywhere $\Delta/E_F\ll 1$ is taken. Dots
correspond to numerically exact results. Solid lines are computed
with an approximation described in Appendix B which becomes exact
for $R/\lambda_F \gg 1$. The inset magnifies results for small
$R$.} \label{Fig:IvsR}
\end{figure}

The most general expression for the current is given in Eq.
(\ref{IvsR}). Below we focus on the limit $\delta,u^{-1}\ll 1$. We
find three regimes of interest, depending on the value of
$R/\lambda_F$.

\subsubsection{Small radius ($R\ll\lambda_F$)}

This limit is not physically realizable, at least with current
materials. However, it is interesting for two reasons. First, it
yields a radius dependence that directly reflects the entangled
nature of the electron current. Second, it can be used as a unit
of current such that, when referred to it, calculated currents
have a range of validity that goes well beyond the geometrical
model here considered. That permits a direct comparison between
different theoretical models and experimental setups.

For $k_F R \ll 1$ we obtain
\begin{equation}
\label{I-inf-R=0} I(R)\simeq I_0\equiv \frac{2\pi}{6^4} J_V \tau^4
k_F^6 R^8.
\end{equation}
This $R^8$ behavior is easy to understand. To compute the current
we must square the matrix element between the initial and the
final state, i.e. the Cooper pair hopping amplitude. The tunneling
of each electron involves an integral over the interface, which
for $k_F R \ll 1$ contributes a factor $R^2$ to the amplitude,
regardless of the incident angle. The Cooper pair amplitude
becomes $\sim R^4$, which leads to the $R^8$ behavior for the
probability.

It is interesting to compare the $R^8$ law here derived with, e.g.
the $R^4$ behavior of the NN tunnel current ($u\gg1$), namely,
\begin{equation} \label{I-NN-inf-R=0}
I_0\simeq\frac{\pi}{9}J_V\tau^2 k_F^2R^4\,,
\end{equation}
or with the $R^6$ dependence for the transmission of photons
through a circular aperture \cite{bethe44}.

Eqs. (\ref{I-inf-R=0}) and Eq. (\ref{I-NN-inf-R=0}) yield the
following relation for the narrow interface conductances:
\begin{equation} \label{relation-G-R-small}
G_{\rm NS}=\frac{h}{4e^2}G^2_{\rm NN} \,\,\, (R\rightarrow 0) \ .
\end{equation}
It is important to note that Eq. (\ref{relation-G-R-small}) still
applies if both conductances are replaced by their
momentum-independent counterparts.

In Fig. \ref{Fig:IvsR} we plot the current density as a function
of the interface radius. Dots represent the exact calculation
taken from Eqs. (\ref{IvsR-delta0}) and (\ref{IvsR-uinf-delta0}),
which we have been able to evaluate numerically for
$u\rightarrow\infty$ (up to $R=1.65\lambda_F$) and $u=10$ (up to
$R=3\lambda_F$), while solid lines are obtained from a
large-radius approximation described in Appendix B. For $u=1.1$
convergence problems prevent us from presenting numerically exact
results.  We find that the small-radius approximation ($\sim R^8$)
is correct within $1 \%$ accuracy up to $R\sim 0.1\lambda_F$.
Above that value it overestimates the current.

\subsubsection{Intermediate radius ($\lambda_F<R<\infty$)}

In this region no analytical expression for the current is
possible. Above $R\approx 2\lambda_F$ even the numerical
calculation of Eq. (\ref{IvsR-uinf-delta0}) (which presumes
$\delta,u^{-1}\ll 1$) is difficult, since for large radii we
cannot compute five strongly oscillating nested integrals. A set
of two approximations which reduces the number of nested integrals
from five to three is discussed in Appendix B and expressed in
Eqs. (\ref{IvsR-approx-delta0}) and (\ref{radio-ojo}).

In Fig. \ref{Fig:IvsR} we plot $I(R)/I(R\rightarrow\infty)$, which
is the total current normalized to the thermodynamic limit
expression (\ref{tunnel-current}) with $A$ in Eq. (\ref{IV})
replaced by $\pi R^2$. For finite barriers, $w=5\lambda_F$ has
been taken. A free parameter has been adjusted to fit the
numerically exact result in the region where it is available. As
explained in Appendix B, such a scheme is particulary well suited
for moderate-to-large radius values. The inset of Fig.
\ref{Fig:IvsR} shows that, as expected, the approximation fails
for small values of $R$, where it yields an $R^4$ behavior instead
of the correct $R^8$ law, thus overestimating the current.

Here we wish to remark that, unlike in the case of a clean NS
point contact \cite{beenakker92,beenakker95}, the radial current
dependence shows {\it no structure of steps and plateaus} as more
channels fit within the area of the interface. This is due the
fact that we operate in the tunneling regime, which decreases the
height of the possible steps and, more importantly, to the
strongly non-adiabatic features of the structure along the
$z$-direction.

\subsubsection{Large radius ($R\rightarrow\infty$)}

While a numerically exact calculation is already nonfeasible for
$R$ a few times $\lambda_F$, the approximation described in
Appendix B becomes increasingly accurate for large $R$. This
allows us to conveniently investigate how the broad interface
limit is recovered [see Eqs. (\ref{tunnel-current}) and
(\ref{ISNu-inf-small-delta})]. Such a limit is characterized by
$I(R)$ growing with  $R^2$, i.e. proportionally to the area,  a
behavior also shown by the NN conductance. Convergence to the
thermodynamic limit is much slower for low barriers than for large
barriers. The reason has to do with {\it focussing}. The wave
length of the characteristic energies $E_{\parallel}=E-E_z$
determines the length scale over which the relative phase between
distant hopping events varies appreciably. This is the distance
over which multiple hopping points (which play the role of
multiple ``Feynman paths") cancel destructively for large radius
interfaces. As discussed in the previous section, low barriers are
more energy selective, making most of the electrons leave with
$E_z$ close to $E_F$ and thus with small $E_{\parallel}$. As a
consequence, saturation to the large radius limit is achieved on
the scale of many times $\lambda_F$. By contrast, high barriers
are less energy selective and give a greater relative weight to
electrons with low $E_z$ and high $E_{\parallel}$. A large
fraction of the electrons has a short parallel wave length. This
explains why, for high barriers, the large $R$ behavior is reached
on a short length scale.


\subsection{Length scales in the thermodynamic limit}

It is known that pairing correlations between electrons decay
exponentially on the scale of the coherence length
$\xi_0=\lambda_F/\pi^2\delta$. This fact is reflected by the
exponential factors contained in the integrands of the equations
for $I(R)$ in Appendix B. Thus one might expect that the
thermodynamic limit relies on such a decay of correlations.

The following argument might seem natural. The double integral
over the interface of area $A$ may be viewed as an integral of the
two-electron center of mass, which yields a factor $A$, and an
integration over the relative coordinate, which is independent of
$A$ due to a convergence factor which expresses the loss of
pairing correlations. The final current would grow as $I \sim
A\xi_0^2\sim A/\delta^2$. However, as discussed in the previous
subsection, the thermodynamic limit is achieved on a much shorter
scale, namely, the Fermi wave length. If an electron leaves
through point ${\bf r}_1$ one may wonder what is the contribution
to the amplitude stemming from the possibility that the second
electron leaves through ${\bf r}_2$, eventually integrating over
${\bf r}_2$. Eq. (\ref{practical-Hamiltonian}) suggests that the
amplitude for two electrons leaving through ${\bf r}_1$ and ${\bf
r}_2$ will involve the sum of many oscillating terms with
different wave lengths, the shortest ones being $\sim \lambda_F$.
This reflects the interference among the many possible momenta
that may be involved in the hopping process. Such an interference
leads to an oscillating amplitude which decays fast on the scale
of $\lambda_F$, rendering the exponentially convergent factor
irrelevant. Thus, in the thermodynamic limit the current tends to
a well defined value for $\xi_0\rightarrow\infty$ ($\delta
\rightarrow 0$). In Appendix B we provide a more mathematical
discussion of this result.

One may also investigate the first correction for small, finite
$\delta$. As indicated in Eqs. (\ref{ISNu-inf}) and
(\ref{ISNu-inf-small-delta}), it increases the current. However,
in the presence of a finite cutoff ($\varrho_c < \infty$), a
nonzero value of $\delta$ generates the opposite trend. As
discussed in Appendix B, at tiny relative distances between
hopping points ($|{\bf r}_1 - {\bf r}_2|\lesssim\delta\lambda_F$),
the amplitude increases considerably. A finite upper momentum
cutoff rounds the physics at short length scales, thus eliminating
such a short-distance increase. The result is that, with a finite
cutoff, the first correction to the $\delta=0$ limit is a
decreasing linear term in $\delta$, as revealed in Eq.
(\ref{ISNu-inf-cut-finite}).

\subsection{Angular distribution and correlation}

\begin{figure}
\includegraphics[width=6cm]{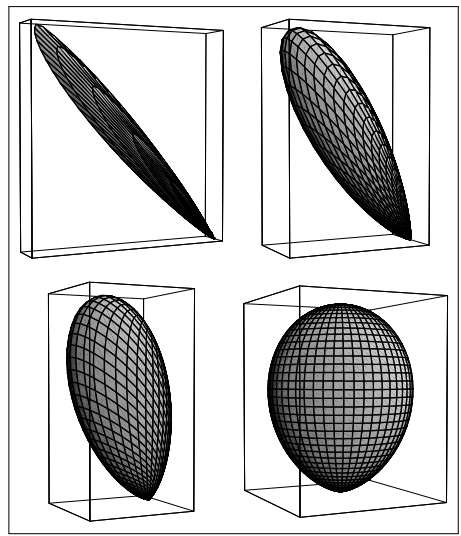}
\caption{Angular correlation profile (in arbitrary units) of the
conditional probability distribution $P(\Omega|\Omega_0)$ that, in
a given tunneling event, an electrons goes into $\Omega$ if the
other electron has gone into $\Omega_0$. Here we plot
$P(\Omega|\Omega_0)$ as a function of $\Omega$ for fixed
$\Omega_0\equiv(\theta_0,\phi_0)=(\pi/4,0)$. From top-left to
bottom-right the radii are: $R=3, 1, 0.5, 0.01\lambda_F$. Observe
that, as $R$ increases, the angular dependence of the second
electron tends to be the conjugate of the first one, i.e. the
distribution becomes peaked around $\Omega=(\pi/4,\pi)$. Note also
that, for small $R$, $P(\Omega|\Omega_0)$ becomes $\propto
\cos^2\theta$ regardless of $\Omega_0$.}\label{3D-2x2}
\end{figure}

We have computed the conditional probability distribution
$P(\Omega|\Omega_0)$ for an electron to be emitted into
$\Omega\equiv (\theta,\phi)$ given that the other electron is
emitted in a fixed direction $\Omega_0$. Such a distribution is
shown in Fig. \ref{3D-2x2} for $\Omega_0=(\pi/4,0)$. We observe
that, for large $R/\lambda_F$, the angular distribution of the
second electron is quite focussed around $\Omega=(\pi/4,\pi)$,
which is mirror-symmetric to $\Omega_0$. As $R/\lambda_F$
decreases, the angular correlation between electrons disappears
and, as a function of $\Omega$, $P(\Omega|\Omega_0)$ becomes
independent of the given value of $\Omega_0$. In particular it
tends to $\sim \cos^2\theta$.

We may also study the probability distribution that one electron
is emitted into direction $\Omega$ regardless of the direction
chosen by the other electron. This amounts to the calculation of
an effective $g(\theta)$ for a finite radius interface to be
introduced in an equation like Eq. (\ref{tunnel-current}) to
compute the current (by symmetry, such a distribution is
independent of $\phi$). As expected, one finds such effective
angular distribution to be $\sim \cos^5\theta$ for large $R$ [see
Eq. (\ref{g-cos-5-theta})], which contrasts with the sharp
$\Omega$-dependence of the conditional angular distribution
$P(\Omega|\Omega_0)$ for given $\Omega_0$.

For small $R$, the effective $g(\theta)$ goes like $\cos^2\theta$,
i.e. it becomes identical to $P(\Omega|\Omega_0)$. This
coincidence reflects the loss of angular correlations. The
$\cos^2\theta$ behavior may be understood physically as stemming
from a random choice of final ${\bf k}_{\parallel}$, which yields
a $\cos\theta$ factor (since $|{\bf k}_{\parallel}|=k_F
\sin\theta$), weighted by a $\cos\theta$ reduction accounting for
the projection of the current over the $z$ direction. An
equivalent study for a NN interface yields also $g(\theta) \propto
\cos^2\theta$. Thus we see that the loss of angular correlations
after transmission through a tiny hole makes the NN and NS
interfaces display similar angular distributions.

The crossover from $g(\theta) \propto \cos^2\theta$ to
$\cos^5\theta$ as $R$ increases involves a decrease of the width
$\Delta\theta$ of the angular distribution. A detailed numerical
analysis confirms this result but reveals that $\Delta\theta$ is
not a monotonically decreasing function of $R$ (not shown).

\section{Nonlocal entanglement in a two-point interface}

Let us turn our attention to a tunneling interface consisting of
two small holes, as depicted in Fig. \ref{SN-Structures}(b). By
``small" we mean satisfying $R/\lambda_F \ll 1$. This is the limit
in which the detailed structure of a given hole is not important
and the joint behavior of the two holes is a sole function of
their relative distance $r$ and the current that would flow
through one of the holes if it were isolated. We expect the
conclusions obtained in this section to be applicable to similar
interfaces made of pairs of different point-like apertures such
as, e.g. two point-contacts or two quantum dots weakly coupled to
both electrodes \cite{recher01}.

The current through a two-point interface has three contributions.
One of them is the sum of the currents that would flow through
each hole in the absence of the other one. Since the two orifices
are assumed to be identical we refer to it as $2I_0$, where $I_0$
is given in Eq. (\ref{I-inf-R=0}). This contribution collects the
events in which the two electrons tunnel through the same opening.
A second contribution $I_e(r)$ comes from those events in which
each electron leaves through a different hole. This is the most
interesting contribution since it involves two non-locally
entangled electrons forming a spin singlet. The third
contribution, $I_{i}(r)$, accounts for the interference between
the previous processes.

\begin{figure}
\includegraphics[width=7cm]{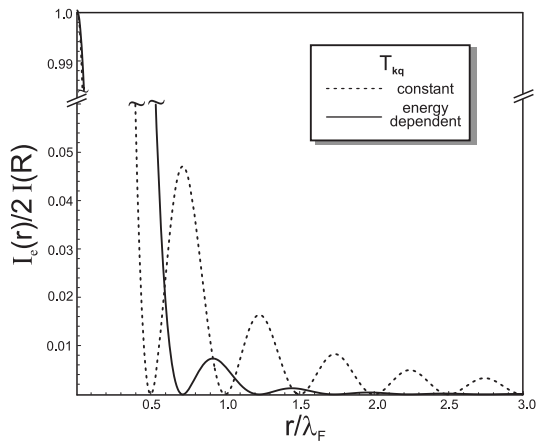}
\caption{Current density through a two-point interface stemming
from non-locally entangled electron pairs, as a function of the
distance between points. The dashed line corresponds to the
current obtained using an energy-independent hopping approximation
whereas in the solid line the correct momentum dependence has been
taken into account.}\label{I2puntos}
\end{figure}

If we write
\begin{equation}
I=2I_0+I_e(r)+I_{i}(r)\,,
\end{equation}
we obtain for the entangled current in the high barrier limit
\begin{equation}
I_e(r)=18I_{0}[B^2(k_Fr)+F^2(k_Fr)B^2(k_Fr)]\,,
\end{equation}
where $B(x)$ in Eq. (\ref{F}) and
\begin{equation} \label{F-x}
F(x)= 3\,\frac{\sin x-x\cos x}{x^3}\,.
\end{equation}
For $\delta \ll 1$, and noting that we are not interested in tiny
distances $r \alt \delta \lambda_F$, we can write
\begin{equation}
\label{I-2p}I_e(r)=2I_{0}[F^2(k_Fr)+F^4(k_Fr)]e^{-2r/\pi\xi_0}\,.
\end{equation}

This is a fast decay because of the geometrical prefactor, which
goes like $r^{-4}$ for $k_Fr \gg 1$. For instance,
$I_e(\xi_0)/I_e(0)\sim 10^{-15}$, with data taken from Al ($\xi_0
\simeq 10^{3} \lambda_F$). For possible comparison with other
tunneling models it is interesting to write the entangled
conductance $G_e(r)\equiv I_e(r)/V$ in terms of the normal
conductance through one narrow hole, $G_{\rm NN}$. Using Eq.
(\ref{relation-G-R-small}), we obtain
\begin{equation} \label{G-ent}
G_e(r)= \frac{h}{2e^2}G_{\rm NN}^2\,
[F^2(k_Fr)+F^4(k_Fr)]e^{-2r/\pi\xi_0}\, .
\end{equation}

To keep track of the interference terms, it is convenient to adopt
a schematic notation whereby $H_T=t_a+t_b$ is the tunneling
Hamiltonian through points $a$ and $b$. Then one notes that, as
obtained from Eqs. (\ref{current}), (\ref{Wfi}) and (\ref{T0}),
the total current can be symbolically written as $I\sim
|(t_a+t_b)(t_a+t_b)|^2$. In this language $I_{0}\sim |t_at_a|^2$.
The $F^2$ term in (\ref{I-2p}) corresponds to $\sim
|t_at_b|^2+|t_bt_a|^2$, while the $F^4$ term stems from the
interference $\sim (t_at_b)(t_bt_a)^*+{\rm c.c.}$ Altogether,
$I_e(r) \sim |t_at_b+t_bt_a|^2$.

The interference current may be divided into two contributions,
\begin{equation} \label{I-interf}
I_{i}=I_{i1}+I_{i2} \ ,
\end{equation}
corresponding to the different types of outgoing channel pairs
which may interfere. The first contribution stems from the
interference between both electrons leaving through point $a$ and
both electrons leaving through point $b$, $I_{i1}\sim
(t_at_a)(t_bt_b)^*+{\rm c.c.}$ One obtains
\begin{equation} \label{interf-1}
I_{i1}(r)=2I_{0}F^2(k_F r) \ .
\end{equation}
$I_{i2}(r)$ comes from the interference between the channel in
which the two electrons leave through the same hole and that in
which they exit through different openings, $I_{i2}(r)\sim
(t_at_a)(t_at_b)^*+{\rm c.c.} $, plus three other equivalent
contributions, altogether summing
\begin{equation}  \label{interf-2}
I_{i2}(r)=8I_{0}F^2(k_F r) e^{-r/\pi\xi_0} \ .
\end{equation}

In the hypothetical case where orifices $a$ and $b$ are connected
to different normal electrodes [e.g. when an opaque barrier
divides into two halves the normal metal of Fig.
\ref{SN-Structures}(b)], the interference contributions
(\ref{interf-1}) and (\ref{interf-2}) would be absent. Then one
would have $I=2I_{0}+I_e(r)$.

\section{Failure of the momentum-independent hopping approximation}

It has been common in the literature on the tunneling Hamiltonian
to assume that the tunneling matrix elements appearing in
(\ref{tunnel-Hamiltonian}) are independent of the perpendicular
momenta $k_zq_z$ (see, for instance, Ref. \cite{mahan00}). Below
we show that, for three-dimensional problems, such an assumption
is unjustified and leads to a number of physical inconsistencies
\cite{comm-cancel}.

For simplicity we focus on the high barrier limit. To investigate
the consequences of the momentum-independent hopping
approximation, we replace Eq. (\ref{hopping-u-inf}) by
\begin{equation} \label{energy-independent}
T_{\mathbf{k}\mathbf{q}}=\frac{\tau}{\Omega N(0)}
\delta(\mathbf{k}_{\parallel}-\mathbf{q}_{\parallel})\,k_{F}^{2}\,,
\end{equation}
i.e. we change $k_zq_z$ by $k_F^2$.

{\it Broad interface}. For a large NS junction, we find that the
total current in units of $I_V$ diverges ($x\equiv\cos\theta$):
\begin{equation}\label{INS-cnst}
I_{\rm{NS}}=\tau^{4}I_{V}\int_{0}^{1}\frac{dx}{x}\:
\frac{x^{2}+\sqrt{x^{4}+\delta^{2}}}{2(x^{4}+\delta^{2})}\rightarrow\infty\,,
\end{equation}
i.e. $I_{\rm{NS}}$ grows faster than $A$ for $A\rightarrow
\infty$. Eq. (\ref{INS-cnst}) is the analogue of Eqs.
(\ref{tunnel-current}) and (\ref{g-high-u}).

A different divergence occurs for a broad NN tunnel junction:
\begin{equation} \label{INN-cnst}
I_{\rm{NN}}=2\tau^{2}I_{V}\int_{0}^{1}\frac{dx}{x}\rightarrow\infty\,,
\end{equation}
which contrasts with the finite integral $I_{\rm NN}\sim
\int_{0}^{1} dx\, x^3$ obtained from inserting
(\ref{g-cos-3-theta}) into (\ref{tunnel-NN-current}).

{\it Local Hamiltonian}. If one attempts to derive the real space
tunneling Hamiltonian with the assumption
(\ref{energy-independent}), one obtains an expression identical to
that in Eq. (\ref{ham-real-space-1}) with $\widetilde{L}(z,z')$
replaced by
\begin{equation}
\label{M-tilda}\widetilde{M}(z,z')=\frac{k_F^2}{L_{z}}
{\sum_{k_{z},q_{z}}}\varphi_{k_z}\!(z)\,\varphi^{*}_{q_z}\!(z')\,.
\end{equation}
As in section IV, we use stationary waves for
$\varphi_{k_z,q_z}(z)$. Invoking the identity
\begin{equation} \label{principal-value}
\sum_{k_z >0}\sin(k_zz)={\cal P}\frac{1}{z}
\end{equation}
we obtain
\begin{widetext}
\begin{eqnarray} \label{ham-principal-value}
H_{T}=\sum_{\sigma}\frac{\tau k_F^2 }{2\pi^{4}N(0)}\int
d\mathbf{r}\int_{-L}^{0}\frac{dz}{z}\int_{0}^{L}\frac{dz'}{z'}
\,\psi^{\dagger}_{N}(\mathbf{r},z;\sigma)
\psi_{S}(\mathbf{r},z';\sigma)+\textrm{H.c.}\,,
\end{eqnarray}
\end{widetext}
where the reference to the principal value has been removed
because, in the tunneling limit, the fields vanish linearly at the
origin.

If we had chosen plane wave functions for $\varphi_{k_z,q_z}$ in
Eq. (\ref{M-tilda}), we would have obtained a different
Hamiltonian, namely,
\begin{equation} \label{si-daga-si}
H_{T}=\sum_{\sigma}\frac{\tau k_{F}^{2}}{8\pi^{2}N(0)}\int\!
d\mathbf{r}\,\psi^{\dagger}_{N}(\mathbf{r},0;\sigma)
\psi_{S}(\mathbf{r},0;\sigma)+\textrm{H.c.}\,,
\end{equation}
which is some times proposed in the literature (see e.g. Ref.
\cite{falci01}). This situation, whereby plane-wave and
stationary-wave representations lead to different, both
unphysical, local Hamiltonians contrasts with the scenario
obtained with the right matrix element. As noted in section IV,
the more physical choice (\ref{hopping-u-inf}) leads in both
representations (plane and stationary waves) to the correct local
Hamiltonian (\ref{Ht-realspace-u-inf}). The fact that Eq.
(\ref{energy-independent}) leads to a wrong real space Hamiltonian
which, moreover, depends on the choice of representation, may be
viewed as further proof of the inadequacy of the
energy-independent hopping model.

{\it Thermodynamic limit}. For a NS interface with
$\delta\rightarrow 0$, a dimensional analysis for
$A\rightarrow\infty$ suggests that the total current $I_{\rm NS}$
diverges non-thermodynamically like $\sim A^2$. For a NN
interface, we find the divergence $A\ln A$.

{\it Unitarity}. The divergences expressed in Eqs.
(\ref{INS-cnst}) and (\ref{INN-cnst}), as well as the related
anomalous thermodynamic behavior, could have been anticipated by
noting that, if $T_{\bf kq}$ is assumed to be independent of
energy, then Eq. (\ref{tunneling-transmission}) must be multiplied
by $(E_F/E_z)^2$. As a result, the transmission probability at
energy $E_z$, which should stay smaller than unity, grows instead
as $T(E_z)\sim E_z^{-1}$ for $E_z \rightarrow 0$. Such a violation
of unitarity necessarily generates a divergent current in the
broad interface limit for both NN and NS interfaces.

{\it Nonlocally entangled current}. Finally, we note that, using
(\ref{energy-independent}), the nonlocally entangled current
through two distant points is
\begin{equation}\label{I'hopp-cnst}
\tilde{I}_e(r)=2\,\tilde{I}_{0}[\tilde{F}^2(k_Fr)+\tilde{F}^4(k_Fr)]e^{-2r/\pi\xi_0}\,,
\end{equation}
where
\begin{equation} \label{F-tilde}
\tilde{F}(x)=\frac{\sin x}{x}\, ,
\end{equation}
with the tildes generally referring to the momentum-independent
approximation. Here, $\tilde{I}_0 = 81 I_0$ is the current through
one narrow hole. Correspondingly, the entangled conductance
$G_e(r)$ is written like in Eq. (\ref{G-ent}) with $F(k_Fr)$
replaced by $\tilde{F}(k_Fr)$.

Comparison of Eqs. (\ref{I-2p}) and (\ref{I'hopp-cnst}) indicates
that the $r$-dependence of the geometrical prefactor is markedly
different: For growing $r$, the nonlocally entangled current
decays much more slowly ($r^{-2}$) than its momentum-dependent
counterpart ($r^{-4}$). It is interesting to compare the ratios
$\lambda(r)\equiv I_e(r)/I_e(0)$ and $\tilde{\lambda}(r)\equiv
\tilde{I}_e(r)/\tilde{I}_e(0)$. While
$\lambda(0)=\tilde{\lambda}(0)=1$ by construction, the ratio
$\lambda/\tilde{\lambda}$ becomes $\sim 6 \times 10^{-4}$ and $2
\times 10^{-7}$ for $r/\lambda_F=20$ and $10^3$, respectively.

{\it Interference terms}. As expected from the comparison of Eqs.
(\ref{I-2p}) and (\ref{I'hopp-cnst}), the interference
contributions are identical to those discussed in the previous
section with $F(k_Fr)$ replaced by $\tilde{F}(k_Fr)$ in Eqs.
(\ref{interf-1}) and (\ref{interf-2}).

{\it Generality of the model}. An important question is whether
our results for the entangled and interference current through
pairs of tiny geometrical holes apply to other, more realistic
pairs of small interfaces such as two point contacts or two
quantum dots \cite{recher01}. The fact that the decay with
distance of the entangled current reported in Refs.
\cite{deutscher00,recher01,falci01,apinyan02,melin02,feinberg02,recher02,feinberg03,recher03}
follows the same law as Eqs. (\ref{I'hopp-cnst}) and
(\ref{F-tilde}) (except for the $\tilde{F}^4$ term there
neglected), suggests that such is indeed the case. Below we prove
this expectation.

Due to Eq. (\ref{practical-Hamiltonian}), the sum in Eq.
(\ref{Tfi}) involves
\begin{equation}
\label{Tfi-explicit}
\sum_{\mathbf{q}}\frac{u_{\mathbf{q}}v_{\mathbf{q}}}
{E_{\mathbf{q}}}q_z^2 e^{-i{\bf q}_{\parallel}\cdot ({\bf
r}_1-{\bf r}_2)} \, .
\end{equation}
This sum over {\bf q} is clearly affected by the presence of the
$q_z^2$ factor, yielding a result $\propto F(k_Fr)$, with $r=|{\bf
r}_1-{\bf r}_2|$. In the momentum-independent hopping
approximation, $q_z^2$ is replaced by $k_F^2$, rendering the sum
$\propto \tilde{F}(k_Fr)$. In fact, the two functions are related:
\begin{equation}
\left. \frac{\partial^2}{\partial z^2}
\tilde{F}(k_F\sqrt{r^2+z^2})\right|_{z=0} = \frac{1}{3}k_F^2 F(k_F
r) \ .
\end{equation}
We note that the distance dependence is determined by the
properties of the superconductor and not by those of the normal
electrode. If a quantum dot mediates between the superconductor
and the normal metal, then an effective hopping must be introduced
in (\ref{Tfi-explicit}) which, however, does not add any new
momentum dependence [see Eq. (11) of Ref. \cite{recher01}].
Departure from the specific type of contact here considered will
translate only into a different value of $I_0$, the distance
dependent prefactor remaining identical. We notice, however, that
the preceding discussion is restricted to the case where
quasiparticle propagation is ballistic in both electrodes, i.e. we
neglect the effect of impurities or additional barriers
\cite{hekk94}.

\section{Summary}

We have investigated the electron current through a NS tunneling
structure in the regime $k_BT \ll eV \ll \Delta$ where Andreev
reflection is the dominant transmissive channel. We have
rigorously established  the physical equivalence between Cooper
pair emission and Andreev reflection of an incident hole. A local
tunneling Hamiltonian has been derived by properly truncating that
of an infinite interface in order to describe tunneling through an
arbitrarily shaped interface. Such a scheme has been applied to
study transport through a circular interface of arbitrary radius
and through an interface made of two tiny holes.
In the former case, the angular correlations between the two
emitted electrons have been elucidated and shown to be lost as the
interface radius becomes small. We have also investigated how the
thermodynamic limit is recovered, showing that, due to the
destructive interference between possible exit points, it is
achieved for radii a few times the Fermi wave length. For the case
of a two-point interface, we have calculated the nonlocally
entangled current stemming from processes in which each electron
leaves the superconductor through a different orifice. We have
found that, as a function of the distance between openings, such
an entangled current decays quickly on the scale of one Fermi wave
length. The interference between the various outgoing two-electron
channels has also been investigated and shown to yield
contributions comparable to the nonlocally entangled current. We
have found that, in a three-dimensional problem, it is important
to employ hopping matrix elements with the right momentum
dependence in order to obtain sound physical results in questions
having to do with the local tunneling Hamiltonian (whose correct
form has also been obtained from a tight-binding description), the
thermodynamic limit, the preservation of unitarity, and the
distance dependence of the nonlocally entangled current through a
two-point interface. An important virtue of the method here
developed is that it enables the systematic study of Cooper pair
emission through arbitrary NS tunneling interfaces and opens the
door to a convenient exploration of the fate of Cooper pairs in
the normal metal and, in particular, to the loss of phase and spin
coherence between emitted electrons.

\acknowledgments

We would like to thank Miguel A. Fern\'andez for useful
discussions. Special thanks are due to Pablo San-Jose for his
great help in some technical and conceptual questions. This work
has been supported by the Direcci\'on General de Investigaci\'on
Cient\'{\i}fica y T\'ecnica under Grant No. BFM2001-0172, the EU
RTN Programme under Contract No. HPRN-CT-2000-00144, and the
Ram\'on Areces Foundation. One of us (E.P.) acknowledges the
support from the FPU Program of the Ministerio de Ciencia y
Tecnolog\'{\i}a and the FPI Program of the Comunidad de Madrid.

\appendix

\section{Discrete vs. continuum space}

Take a discrete chain made of $N$ sites with period $a$ described
by the Hamiltonian
\begin{equation}
H_0=-t\sum_{i=1}^{N-1}c_{i+1}^\dagger c_i+\rm{H.c.}\,,
\end{equation}
where $t=\hbar^2/2ma^2>0$ is the hopping parameter that yields an
effective mass $m$ in the continuum limit.

The eigenstates of this chain are of the form
\begin{equation}
|\phi_n\rangle=\left ( \frac{2}{N+1}\right )
^{1/2}\sum_{i=1}^{N}\sin(k_nz_i)c^{\dagger}_i|{\rm vac} \rangle\,,
\end{equation}
where $z_i=i a$ and $k_n=\pi n/a(N+1)$ with $i,n\in[1,N]$. The
eigenvalues are
\begin{equation}
E_n=-2t\cos(k_na)\,.
\end{equation}
The basis set $\{|\phi_n\rangle\}$ is orthonormal. Thus we may
write
\begin{eqnarray}
c^{\dagger}_{k_n}&=&\left ( \frac{2}{N+1}\right )
^{1/2}\sum_{i=1}^{N}\sin(k_n z_i)c^{\dagger}_i\,,\\
c^{\dagger}_{i}&=&\left ( \frac{2}{N+1}\right )
^{1/2}\sum_{n=1}^{N} \sin(k_n z_i)c^{\dagger}_{k_n}\,.
\end{eqnarray}

We write the transfer Hamiltonian between two $N$-site chains as
\begin{eqnarray}
H_T&=&-t' a^{\dagger}_1 b_{-1}+\rm{H.c}\label{Ht-discreet-0}\\
&=&\frac{2t'}{N+1}\sum_{n=1}^{N}\sum_{m=1}^{N}\sin(k_na)
\sin(k_ma) a^{\dagger}_{k_n}b_{k_m}+\rm{H.c.}~~~~~~
\label{Ht-discreet}
\end{eqnarray}
which may be treated as a small perturbation when $t'\ll t$.

To investigate the continuum limit, we take $a\rightarrow0$ and
$t\rightarrow \infty$ so that $m$ and $k_F$ remain finite. We also
take $N\gg1$. Noting that the sine functions in
(\ref{Ht-discreet}) can be approximated by their arguments $k_na
\lesssim k_Fa\ll 1$ and that
$k_{\rm{max}}=\pi/a\rightarrow\infty$, we get
\begin{equation}\label{bilinear}
H_T=\frac{2t'a^3}{L}\sum_{k,q>0}k\:q\,a^{\dagger}_k b_q+\rm{H.c.}
\end{equation}
This Hamiltonian is bilinear in the momenta of the electron on the
right and left chain. If we were in 3D we would specify that the
bilinearity refers to the momenta perpendicular to the interface
plane. This hamiltonian is analogous to that which we proposed for
the continuum Bardeen theory in the case of a high barrier [see
Eqs. (\ref{tunnel-Hamiltonian}) and (\ref{hopping-u-inf})].

We may work out the corresponding Hamiltonian in real space. For
that we note that, in the continuum limit, $H_T$ in Eq.
(\ref{Ht-discreet-0}) can be expressed in terms of field operators
evaluated in $z=\pm a$. When $a\rightarrow0$, the field operators
can be expanded as
\begin{equation}
\psi(a)=\psi(0)+a\left.\frac{d\psi(z)}{dz}\right|_{z=0}+{\cal
O}(a^2)\,,
\end{equation}
where $\psi(0)=0$ is a condition that results naturally from the
properties of the wave functions in a chain starting in $i=1$ or
$i=-1$. For such chains, $i=0$ is an imaginary point where the
wave function necessarily vanishes; it is the place where we would
locate the hard wall in a continuum description \cite{sols89}.
Then the tunneling Hamiltonian can be written
\begin{equation} \label{derivatives}
H_T=t'a^3\left.\frac{d\psi^{\dagger}_{R}(z)}{d z}\right|_{z=0}
\left.\frac{d\psi_{L}(z')}{d z'}\right|_{z'=0}+\textrm{H.c.}
\end{equation}
This Hamiltonian is exactly the one-dimensional version of that in
Eq. (\ref{Ht-realspace-u-inf}). The fact that we have derived it
from a completely different set of physical arguments should be
viewed as a definite proof of the adequacy of the tunneling
Hamiltonians proposed in section IV. The Hamiltonians
(\ref{bilinear}) and (\ref{derivatives}) have been obtained in the
continuum limit. On the other hand, Eqs. (\ref{hopping-u-inf}) and
(\ref{Ht-realspace-u-inf}) were derived for high barriers or,
equivalently, low energies. Clearly, this is not a coincidence,
since it is at low energies where the long wavelengths make the
electron move in the chain as in continuum space.

\section{Total current vs. interface radius}

To calculate the total current as a function of the interface
radius $R$ we have to evaluate the matrix element (\ref{Tfi})
using hopping energies obtained from the tunneling Hamitonian
(\ref{practical-Hamiltonian}). In the resulting expression we need
to integrate over the final momenta of the two electrons in the
normal metal, the momentum of the intermediate virtual state
consisting of a quasiparticle in the superconductor, as well as
the coordinates of the points where each electron crosses the
interface area. The integrations over the momenta in the final
state lead to four angular integrals
($\theta_{1,2}\in[0,\pi/2];\varphi_{1,2}\in[0,2\pi]$), the moduli
being fixed by the condition $k_BT,eV\rightarrow 0$. The
integration over the superconductor excited states leads to three
integrals: $\theta_{s}\in[0,\pi], \varphi_{s}\in[0,2\pi],
q\in[0,q_c]$. On the other hand, integration over the hopping
points of each electron leads to two interface integrals
($r_{1,2}\in[0,R]$, $\theta_{r1,r2}\in[0,2\pi]$), which makes four
more integrals, totalling eleven real variables to be integrated.
Using the symmetry property that the integrand is independent of
one azimuthal angle, and  solving analytically the four real space
integrals, we are left with six non-reducible nested integrals of
strongly oscillating functions.

We define $\vec{\kappa}\equiv\mathbf{k}/k_F$,
$\vec{\varrho}\equiv\mathbf{q}/k_F$. Since the modula of the final
momenta are fixed by conservation requirements, we may write
$\kappa_{\parallel i}=\sin\theta_i$, $\kappa_{zi}=\cos\theta_i$
($i=1,2$). For the virtual states in the superconductor:
$\varrho_{\parallel}=\varrho\sin\theta_s$,
$\varrho_z=\varrho\cos\theta_s$.

The general, exact formula for the total current as a function or
$R$ is
\begin{widetext}
\begin{eqnarray}
\label{IvsR} I(R)&=&I_V\tau^4\frac{(k_FR)^2}{4\pi^3}\int
d\Omega_1\int d\kappa_{2z}
\kappa_{1z}^2\kappa_{2z}^2e^{2p_0w[1-b(\kappa_{1z},\kappa_{2z})]}\nonumber\\
&\times&\left[\frac{2}{\pi}\int
d\varrho\frac{\delta}{(\varrho^2-1)^2+\delta^2} \varrho^4\int
d\Omega_se^{p_0w[1-b(\varrho_z)]}\prod_{j=1,2}a(\kappa_{zj},\varrho_z)
J(|\vec{\varrho}_\parallel -
\vec{\kappa}_{1\parallel}|,|\vec{\varrho}_\parallel +
\vec{\kappa}_{2\parallel}|)\right]^2,\nonumber\\
\end{eqnarray}
where $J$ is a short-hand notation for
\begin{equation}\label{N}
J(x,y)\equiv \frac{J_1(k_FR\,x) J_1(k_FR\,y)} {x\,y}\,.
\end{equation}
The first-order Bessel functions result from the exact integration
over the tunneling points ${\bf r}_1$ and ${\bf r}_2$.

For $\delta\rightarrow0$, the Lorentzian becomes a delta function
and the integral over $\varrho$ is evaluated exactly. We get (with
$u$ still arbitrary)
\begin{eqnarray} \label{IvsR-delta0}
I(R)&=&I_V\tau^4\frac{(k_FR)^2}{4\pi^3}\int d\Omega_1\int
d\kappa_{2z}
\kappa_{1z}^2\kappa_{2z}^2e^{2p_0w[1-a(\kappa_{1z},\kappa_{2z})]}\nonumber\\
&\times&\left[\int d\Omega_s e^{p_0w[1-b(\varrho_z)]}
\prod_{j=1,2}a(\kappa_{zj},\varrho_z) J(|\vec{\varrho}_\parallel -
\vec{\kappa}_{1\parallel}|,|\vec{\varrho}_\parallel +
\vec{\kappa}_{2\parallel}|)\right]^2.
\end{eqnarray}

For $\delta$ arbitrary and $u\gg1$, Eq. (\ref{IvsR}) becomes
\begin{equation}
\label{IvsR-uinf} I(R)=I_V\tau^4\frac{(k_FR)^2}{4\pi^3}\int
d\Omega_1\int
d\kappa_{2z}\kappa_{1z}^2\kappa_{2z}^2\left[\frac{2}{\pi}\int
d\varrho\frac{\delta}{(\varrho^2-1)^2+\delta^2} \varrho^4\int
d\Omega_s J(|\vec{\varrho}_\parallel -
\vec{\kappa}_{1\parallel}|,|\vec{\varrho}_\parallel +
\vec{\kappa}_{2\parallel}|)\right]^2.
\end{equation}
Finally, for both $\delta\rightarrow0$ and $u\gg1$, we obtain
\begin{equation}
\label{IvsR-uinf-delta0} I(R)=
I_V\tau^4\frac{(k_FR)^2}{4\pi^3}\int d\Omega_1\int
d\kappa_{2z}\kappa_{1z}^2\kappa_{2z}^2\left[\int
d\Omega_sJ(|\vec{\varrho}_\parallel -
\vec{\kappa}_{1\parallel}|,|\vec{\varrho}_\parallel +
\vec{\kappa}_{2\parallel}|)\right]^2,
\end{equation}
\end{widetext}
which for $k_FR \ll 1$ leads to Eq. (\ref{I-inf-R=0}) in the main
text. This is easy to see considering that
$\lim_{x\rightarrow0}J_1(k_FR\,x)/x=k_FR/2$.

Even after making $\delta,u^{-1} \rightarrow 0$, the resulting
expression (\ref{IvsR-uinf-delta0}) is such that a numerical
integration for arbitrary $R$ is not yet possible. In order to
evaluate (\ref{IvsR-delta0}) and (\ref{IvsR-uinf-delta0})
numerically we need to introduce a set of {\it two approximations}
which are good for $k_FR\gg 1$ and reasonable for intermediate
$R$. To introduce the {\it first} approximation we go back to the
original expression (\ref{current}), where the space coordinates
have not yet been integrated.  Then we shift from the two space
coordinates $(\mathbf{r}_1,\mathbf{r}_2)$ to center-of-mass and
relative coordinates $(\mathbf{r}_c,\mathbf{r})$. The integration
domain of the center-of-mass coordinate $\mathbf{r}_c$ is still a
circle of radius $R$. However, the integration region of the
relative coordinate $\mathbf{r}$ is more complicated: It is
eye-shaped and centered around $\mathbf{r}_c$. The first
approximation consists in assuming that, for all $\mathbf{r}_c$,
the integration domain of the relative coordinate is circular
instead of eye-shaped. The area of such a circular region is a
free parameter which can be adjusted by, e.g. comparing the
approximate result with the exact calculation for those values of
$R$ for which the latter can be performed.

It is intuitive (and rigorously proved in subsection VII.B) that,
because of diffraction, when $R\alt\lambda_F$, the parallel
momentum is not conserved and, in particular, the two electrons do
not leave necessarily with opposite parallel momenta [see Fig.
(\ref{3D-2x2})]. Nevertheless, as $R$ increases the interface
begins to be large enough so as to permit parallel momentum to
become better conserved. A quasi-delta function
$\widetilde{\delta}(\mathbf{k}_{1\parallel}+\mathbf{k}_{2\parallel})$
effectively appears. In particular we have:
$\lim_{R\rightarrow\infty}J_1(k_{\parallel}
R)/k_\parallel=2\pi\delta(\mathbf{k}_\parallel)/R$. Thus, our {\it
second} approximation consists in assuming that, for all
$R>\lambda_F$, the quasi-delta is an exact delta:
$\widetilde{\delta}\rightarrow\delta$. This is equivalent to the
assumption that there is no diffraction, i.e. that we work in the
ray optics limit. This approximation becomes exact as
$R\rightarrow\infty$ and it is a reasonable one for finite radii.
Of course, this approximation fails for $R\alt\lambda_F$, yielding
a wrong $R^4$ behavior.

With the two previous approximations we can reduce the number of
numerical integrals from five to three. To write the resulting
expressions, let us introduce some compact notation. We define
$x\equiv\cos\theta$ (where $\theta$ is the angle formed by the
outgoing momentum with the direction normal to the interface),
$y\equiv\cos\theta_q$ ($\theta_q$ having a similar definition
within the superconductor), $\lambda\equiv k_F|\mathbf{r}_c|$, and
$\mu\equiv k_F|\mathbf{r}|$.

For $\delta\rightarrow0$ and arbitrary $u$ we obtain
\begin{widetext}
\begin{eqnarray}
\label{IvsR-approx-delta0}I(R)\simeq I_V\tau^4
\int_0^1dx\:x^3e^{2p_0w[1-b(x)]}\left\{
\int_0^{k_FR}d\lambda\:\frac{2\lambda}{(k_FR)^2}\int_0^1dy\frac{r(\lambda)y^2}{x^2-y^2}
[a(x,y)]^2e^{2p_0w[1-b(y)]}\right.\nonumber\\
\left.\times\left[\sqrt{1-y^2}\;J_0\!\!\left(r(\lambda)\sqrt{1-x^2}\right)
J_1\!\!\left(r(\lambda)\sqrt{1-y^2}\right)-
\sqrt{1-x^2}\;J_0\!\!\left(r(\lambda)\sqrt{1-y^2}\right)
J_1\!\!\left(r(\lambda)\sqrt{1-x^2}\right)\right]\right\}^2\,,\nonumber\\
\end{eqnarray}
\end{widetext}
where $r(\lambda)$ is the radius of the approximate circular
domain over which the relative coordinate ${\bf r}$ is integrated.
If the circle is assumed to have the same area as the eye, we
obtain
\begin{equation} \label{exact-area}
r(\lambda)\equiv \sqrt{\frac{8}{\pi}}\left[(k_FR)^2\arccos
\left(\frac{\lambda}{k_FR}\right)-\lambda\sqrt{(k_FR)^2-\lambda^2}\right]^{1/2},
\end{equation}
but in practice this criterion is found to overestimate the total
current. Thus we decide to adopt the ansatz
\begin{equation}
\label{radio-ojo}r(\lambda)\equiv
2k_FR\left(1-\frac{\lambda}{k_FR}\right)^\alpha\,,
\end{equation}
where $\alpha$ is a parameter to be adjusted by comparison with
the exact solution in those cases where it can be computed. In
particular, $\alpha$ has been adjusted from the last two exact
numerical values of each curve, i.e. from the two largest
computationally possible radii. We note that both
(\ref{exact-area}) and (\ref{radio-ojo}) satisfy the requirement
$r(\lambda) \rightarrow 2k_FR$ for $\lambda \rightarrow 0$. The
value $\alpha=1$ corresponds to the case where the circle is
chosen to be the maximum circle which fits within the eye-shaped
integration domain. As expected, this criterion underestimates the
current. The formula (\ref{exact-area}), which overestimates the
result, can be approximated with $\alpha \approx 0.7$. Thus it
comes as no surprise that the value of $\alpha$ obtained by
comparing with the exact result (when available) is an
intermediate number, namely, $\alpha=0.84$, which has been used
for the NS curves in Fig. \ref{Fig:IvsR}.

For arbitrary $\delta$ and $u\gg 1$, the total current becomes
\begin{widetext}
\begin{eqnarray}
\label{IvsR-approx-uinf}I(R)&\simeq& I_V\tau^4\int_0^{1}dx\:x^3
\left[\frac{2}{(k_FR)^2}\int_0^{k_FR}d\lambda\:\lambda\int_0^{r(\lambda)}d\mu\:\mu
J_0(\mu\sqrt{1-x^2})B(\mu)\right]^2\,,\\
\label{F}B(\mu)&=&\left\{\frac{\sin[S(\delta)\mu)]}{\mu^3}-
\frac{\sqrt[4]{1+\delta^2}\cos[\arctan\delta/2+S(\delta)\mu]}{\mu^2}\right\}
e^{-D(\delta)\mu}\,,
\end{eqnarray}
\end{widetext}
where
\begin{eqnarray}
S(\delta)=\left (\frac{\sqrt{1+\delta^2}+1}{2}\right )^{1/2}
\stackrel {\delta\ll1}{\rightarrow}1\,,\\
D(\delta)=\left (\frac{\sqrt{1+\delta^2}-1}{2}\right )^{1/2}
\stackrel{\delta\ll1}{\simeq}\frac{\delta}{2}\,.
\end{eqnarray}
Thus, for $\delta \ll 1$ we may write
\begin{equation} \label{F-mu-aprox}
B(\mu)\simeq\left[\frac{\sin(\mu)}{\mu^3}-
\frac{\cos(\mu+\delta/2)}{\mu^2}\right] e^{-\delta\mu/2}\,.
\end{equation}

The effect of the phase-shift $\delta/2$ is only appreciable for
$\mu \alt \delta$, i.e. for $r \alt \delta \lambda_F \ll
\lambda_F$, as can be seen by expanding $B(\mu)$ for small $\mu$:
\begin{equation}
B(\mu)=\frac{\delta}{2\mu}+\frac{1}{3}-\frac{\delta\mu}{4}
-\frac{\delta^2}{8}+{\cal O}(\delta^4,\mu^2).
\end{equation}
The phase-shift generates a divergence for $\mu \rightarrow 0$.
Although integrable thanks to the multiplying $\mu$ factor in Eq.
(\ref{IvsR-approx-uinf}), this divergence affects the final
result. Its range of relevance may be estimated by making
$\delta/2\mu$ equal to the limiting value $1/3$ which one would
obtain with $\delta=0$. This yields a range
$r_0=(3/4\pi)\delta\lambda_F$, which will be washed out by any
realistic momentum cutoff $q_c \sim k_F \ll k_F/\delta$.

Finally, we note that comparison of Eqs. (\ref{I-2p}) and
(\ref{F-mu-aprox}) clearly reveals that the entangled current
$I_e(r)$ given in (\ref{I-2p}) is essentially proportional to
$B^2(k_Fr)$. As discussed in Sec. IX, $I_e(r)$ decays faster than
the prefactor obtained from momentum-independent hopping matrix
elements [see Eq. (\ref{I'hopp-cnst})]. The current increase which
results from such an unphysical approximation translates into a
divergent thermodynamic limit (see also Sec. IX).


\begin{references}
\bibitem{com-duke} The low-energy linear dependence $T(E_z)\sim E_z$ and the
related bilinear dependence $T_{\mathbf{k}\mathbf{q}}\sim k_z q_z$
is implicit in Ref. \cite{duke69}. We note, however, that here we
find perfect agreement between Bardeen's perturbative method and
the exact results in the tunneling limit.
\bibitem{comm-morten99} A study of the angular dependence of
Andreev reflection in the broad interface limit has been presented
in Ref. \cite{mortensen99}. However, it is restricted to a delta
barrier interface and consider the case where $N$ is a doped
semiconductor.
\bibitem{comm1} There, under the constraint of a given total
current, entropy is maximized by the electron system adopting a
displaced Fermi sphere configuration which exactly yields the
$\cos\theta$ law.
\bibitem{comm-BTK} Together with the factor appearing in the
definition (\ref{IV}), this factor $1/2$ yields the $1/4$
geometrical correction given by Ref. \cite{sharvin65} and cited in
Ref. \cite{blonder82}. We emphasize however that such $1/4$
correction applies only to the normal conductance, as derived in
Ref. \cite{sharvin65}, but not to the NS interface where, rather,
the correct geometrical correction is $1/12$, as implicitly noted
in Eqs. (\ref{IV}) and (\ref{ISNu-inf-cut-finite}).
\bibitem{comm-cancel} We do not rule out, however, that the errors
derived from the use of Eq. (\ref{energy-independent}) may cancel
in the calculation of some physical quantities such as e.g. the
ratio between the critical current and the normal conductance of a
superconducting tunnel junction \cite{ambegaokar63,mahan00}.
\end{references}
\end{document}